\documentclass[superscriptaddress,aps,twocolumn,a4paper,showpacs]{revtex4}
\usepackage[latin1]{inputenc}
\usepackage[english]{babel}
\usepackage{amsmath}
\usepackage{amsfonts}
\usepackage{amssymb}
\usepackage{graphicx}

\usepackage[colorlinks=true, pdfstartview=FitV, linkcolor=blue, citecolor=red, urlcolor=magenta, breaklinks=true]{hyperref}

\usepackage[all]{xy}
\usepackage{amssymb}
\usepackage{color}
\usepackage{subeqnarray}
\usepackage{subfigure}
\usepackage{epstopdf}

\newcommand{\bes}{\begin{subequations}}
\newcommand{\ees}{\end{subequations}}
\def\ben{\begin{eqnarray}}
\def\een{\end{eqnarray}}
\newcommand{\bens}{\begin{subeqnarray}}
\newcommand{\eens}{\end{subeqnarray}}
\def\be{\begin{equation}}
\def\ee{\end{equation}}

\begin{document}

\title{Compact Vortices}
\author{D. Bazeia}\email{bazeia@fisica.ufpb.br}\affiliation{Departamento de F\'\i sica, Universidade Federal da Para\'\i ba, 58051-970 Jo\~ao Pessoa PB, Brazil} 
\author{L. Losano}\email{losano@fisica.ufpb.br}\affiliation{Departamento de F\'\i sica, Universidade Federal da Para\'\i ba, 58051-970 Jo\~ao Pessoa PB, Brazil}
\author{M.A. Marques}\email{mam.matheus@gmail.com}\affiliation{Departamento de F\'\i sica, Universidade Federal da Para\'\i ba, 58051-970 Jo\~ao Pessoa PB, Brazil} 
\author{R. Menezes}\email{rmenezes@dce.ufpb.br}\affiliation{Departamento de Ci\^encias Exatas, Universidade Federal da Para\'\i ba, 58297-000 Rio Tinto, PB, Brazil}\affiliation{Departamento de F\'\i sica, Universidade Federal de Campina Grande, 58109-970 Campina Grande, PB, Brazil}
\author{I. Zafalan}\email{ivzafalan@gmail.com}\affiliation{Departamento de F\'\i sica, Universidade Federal da Para\'\i ba, 58051-970 Jo\~ao Pessoa PB, Brazil} 
\date{\today}
\begin{abstract}
We study a family of Maxwell-Higgs models, described by the inclusion of a function of the scalar field that represent generalized magnetic permeability. We search for vortex configurations which obey first-order differential equations that solve the equations of motion. We first deal with the asymptotic behavior of the field configurations, and then implement a numerical study of the solutions, the energy density and the magnetic field. We work with the generalized permeability having distinct profiles, giving rise to new models, and we investigate how the vortices behave, compared with the solutions of the corresponding standard models. In particular, we show how to build compact vortices, that is, vortex solutions with the energy density and magnetic field vanishing outside a compact region of the plane.
\end{abstract}
\maketitle

%%%%%%%%%%%%%%%%%%%%%%%
\section{Introduction}

This work deals with vortices in generalized Maxwell-Higgs model in the three-dimensional spacetime. As it is well-known, vortices are planar structures of topological nature \cite{3}, and their importance in high energy physics can be found, for instance, in Refs.~\cite{Vilenkin,Manton}. In particular, they may appear in a phase transition during the cosmic evolution of our Universe \cite{Vilenkin}. They are also of current interest to other areas of Physics; in condensed matter, they may appear in superconductors, and may also be present as magnetic domains in magnetic materials \cite{mag}   

The generalized Maxwell-Higgs model in which we are interested appeared in the beginning of the nineties, with focus on the presence of vortex solutions \cite{1}-\cite{2}. The model includes a function $G(|\phi|)$ of the Higgs field multiplying the Maxwell term, and for a very specific choice of this function, the generalized system supports solutions that map the vortices of the Chern-Simons-Higgs system \cite{4}-\cite{6}. The difference here is that the vortices are electrically neutral, although the magnetic flux exists and is quantized. The function $G(|\phi|)$ can be seen as a kind of generalized {\it magnetic permeability}, and the limit $G \to 1$ leads us back to the standard Maxwell-Higgs model. 

In this work we study the generalized model under specific circumstances, considering several new possibilities. One starts in Sec~\ref{sec1}, reviewing the standard Maxwell-Higgs system and introducing the generalized model, with focus on the first-order formalism which we use to describe explicit solutions of the Bogomol'nyi-Prasad-Sommerfield (BPS) type \cite{7,7a}; see also Ref.~\cite{Vega}. We then investigate two new models in Sec.~\ref{sec3}, and in Sec.~\ref{sec4} we investigate models that allow for the presence of compact vortices, that is, for vortex-like solutions which engender energy density and magnetic field that vanish outside a compact interval of the radial coordinate. 

The motivation to study compact vortices comes from the recent advances in the study and manipulation of materials at the nanometric scale. For instance, in Ref.~\cite{DW} it was experimentally observed that domain walls may modify conformation in constrained geometries, so one can also ask if the miniaturization of magnetic materials can modify the conformational structure of vortices and skyrmions \cite{MS1,MS2}, as in the case recently investigated in \cite{MS3}. In this sense, it seems of current interest to study the possibility of shrinking topological objects such as vortices to compact regions. The study of compact vortices is also part of the recent work on compact structures, such as kinks and lumps \cite{CK,db1}, and Q-balls \cite{db3}. These investigations are based on distinct mechanisms, and the results show that there is no obvious way to make vortices shrink to a compact region of the plane. Here, however, we follow the route proposed in \cite{CK} and show how to construct compact vortices in the generalized model of the Maxwell-Higgs type.

%%%%%%%%%%%%%%%%%%%%%%%%%%%%%%%%%%%%%%%%%%%%
\section{The Model}\label{sec1}

Vortices are topological structures that appear in the three-dimensional spacetime. The Lagrange density that describes the standard Maxwell-Higgs model has the form
\begin{equation}\label{eqm1}
{\cal L}={-\frac{1}{4}}F_{\mu\nu}F^{\mu\nu}+|D_{\mu}\phi|^2-V(|\phi|)\,,
\end{equation}\\
where $F_{\mu\nu}=\partial_\mu A_\nu-\partial_\nu A_\mu$ is the electromagnetic field strength, $A_\mu$ is the electromagnetic vector potential, $\phi$ is the complex scalar field, $D_{\mu}=\partial_{\mu}+ieA_{\mu}$ is the covariant derivative, $e$ is the electric charge and $V(|\phi|)$ is the potential for the scalar field. We are working in the $(2,1)$ dimensional space-time with Minkowski metric $\eta_{\mu\nu}$, with diagonal elements $(1,-1,-1)$. We are also using natural units such that $\hbar=c=1$. In the standard case, the Higgs potential has the form
\begin{equation}
V(\phi)=\frac12 \lambda (v^2-|\phi|^2)^2
\end{equation}
where $\lambda$ is real and positive parameter that describes the strength of the field self-interaction, and $v$ is another real and positive parameter that sets the scale of spontaneous symmetry breaking. 

Before one moves on and introduces the new model, it is of interest to know some specific features of the standard model, in particular the dimension of the several quantities that appear in the model. Since one is working with $(2,1)$ spacetime dimensions, one notes that the field $A_\mu$ has dimension of energy to the power $1/2$ or, in short, dim$(A_\mu)=1/2$. Thus, the other quantities obey: dim$(\phi)=1/2$, dim$(e)={\rm dim}(v)=1/2$, and dim$(\lambda)=1$. The model engenders spontaneous symmetry breaking and it is also known to support vortex solutions, as first studied in \cite{3} and later in \cite{7,Vega}, with focus on the presence of solutions that solve first-order differential equations. 

In order to introduce the new model, we modify the above scenario and consider the Lagrange density
\begin{equation}\label{eqm1}
{\cal L}={-\frac{1}{4}}G(|\phi|)F_{\mu\nu}F^{\mu\nu}+|D_{\mu}\phi|^2-V(|\phi|)\,,
\end{equation}\\
where $G(|\phi|)$ is a dimensionless function of the scalar field. This modification can be seen to describe the presence of a generalized magnetic permeability. It appeared before in \cite{1,2} in the context of vortex solutions, and here we use it again, motivated to describe new models and solutions, with focus on the presence of compact vortex. This model was also considered in \cite{W} in the study of the gauge embedding procedure that produces the dual mapping of the self-dual vector field theory into a Maxwell-Chern-Simons system. More recently, further additions have been considered to describe planar and spatial structures in generalized scenarios; see, e.g., Refs.~\cite{Hora2,Ram} and references therein.

%%%%%%%%%%%%%%%%%%%%%%%%%%%%%%%%%%%%%%%%%%%%
\subsection{Basic considerations}\label{sec2}

The equations of motion of the generalized model \eqref{eqm1} have the form
\bes
\begin{eqnarray}\label{eqm}
D_{\mu}D^{\mu}\phi+{\frac{1}{4}{\frac{\partial{G}}
{\partial{\bar{\phi}}}}F_{\mu\nu}F^{\mu\nu}}+{\frac{\partial{V}}{\partial{\bar{\phi}}}} &=& 0 \label{eqm2}, \\
{\partial_{\mu}{(GF^{\mu\nu})}}+J^{\nu} &=& 0 \label{eqm3}\,,
\end{eqnarray}
\ees
where $J^\mu$ is the conserved Noether current, which is given by $J^{\mu}=-ie[\bar{\phi}D^{\mu}{\phi}-{\phi{\overline{D^{\mu}{\phi}}}}].$
Also, the energy-momentum tensor takes the form
\begin{equation}\label{tem}
T_{\mu\nu}= G(|\phi|)F_{\mu\lambda}F^{\lambda}_{\nu}+{\overline{D_{\mu}{\phi}}}{D_{\nu}{\phi}}+{\overline{D_{\nu}{\phi}}}{D_{\mu}{\phi}}-g_{\mu\nu}{\cal L}\,.
\end{equation}

To search for vortices, one supposes that the fields are all static. As a consequence, if one takes the temporal gauge, $A_{0}=0$, one sees that the electric field vanishes, so the vortex solutions are then electrically neutral. Note that the choice $A_0=0$ is compatible with the $\nu=0$ component of the equation of motion \eqref{eqm3}. Also, the only non-vanishing component of the magnetic field is $B=F^{21}=-F^{12}$.

The energy density is given by
\begin{equation}\label{energy}
{\large{\varepsilon}} =T_{00}= {\frac{1}{2}}G(|\phi|)B^2+|D_{i}{\phi}|^2+V(|\phi|)\,.
\end{equation}
We suppose that the field configurations have the form
\bes\label{anz}\ben
\phi &=& vg(r)e^{in\theta}\label{anz1}, \\
\vec{A} &=& -{\frac{\hat{\theta}}{er}[a(r)-n]}\label{anz2}\,,
\een\ees
where $r$ and $\theta$ are the radial and angular coordinates, respectively, with $r \in [0,\infty)$ and $\theta \in [0,2\pi)$. Also, $n$ is a nonvanishing integer, the vorticity or winding number; $n=\pm1,\pm2,\cdots$. It counts how many times the scalar field winds around itself as
$\theta$ varies in the interval $[0,2\pi)$.

In order to avoid singularities and have finite energy, the field configurations must obey
\bes
\begin{eqnarray}
a(0) &=& n,\quad \quad \quad  g(0) = 0\,, \\
\displaystyle\lim_{r\to \infty} a(r) &=& 0\,, \quad \displaystyle\lim_{r\to \infty} g(r) = 1\,.
\end{eqnarray}
\ees
Now, using the ansatz given by Eqs. \eqref{anz}, we can rewrite the equations of motion as
\bes \label{eqm}
\begin{eqnarray}\label{eqm1a}
\frac{1}{r}[rg']'-\frac{a^2g}{r^2}-\frac{1}{2v^2}V_{g}-\frac{1}{4e^2v^2}G_{g}{\left[\frac{a'}{r}\right]}^2\!\! &=& 0\,, \\ 
rG{\left[\frac{a'}{r}\right]}'-{2e^2v^2g^2a}+a'g'G_g &=& 0\,, \label{eqm2a}
\end{eqnarray}
\ees
with the prime denoting differentiation with respect to the radial coordinate $r$, and $G_g=dG/dg$.
Also, for the field configurations given by Eqs. \eqref{anz}, the magnetic field becomes
\begin{equation}\label{B}
B=-{\frac{a'}{er}}\,.
\end{equation}
Moreover, the angular momentum is given by
\be
J=\int{d^2r\epsilon_{ij}x_{i}T_{0j}},
\ee
and since $T_{0i}=0$, the solutions have vanishing angular momentum. 

The presence of the magnetic field $B$ allows that we introduce the magnetic flux, which has the form
\be 
\Phi=2\pi\int_0^{\infty} r\, dr\,B(r).
\ee
If one uses Eq.~\eqref{B}, it follows that 
\be
\Phi={\frac{2\pi}{e}n}\,,
\ee
The magnetic flux is then a conserved quantity, the topological invariant that only takes multiple values of the basic flux $2\pi/e$. 

If one uses \eqref{energy}, for static fields the energy density can be written in the form
\begin{eqnarray}\label{deny1}
\large{\varepsilon} &=& |(D_{1}\pm iD_{2}){\phi}|^2+{\frac{G(|\phi|)}{2}}{\left[ B \pm {\frac{e(|\phi|^2 - v^2)}{G(|\phi|)}} \right]^2}\nonumber\\
&-& {\frac{e^2}{2}}{\frac{(|\phi|^2-v^2)^2}{G(|\phi|)}}+V(|\phi|)\pm ev^2B.
\end{eqnarray}
Thus, if one imposes that
\begin{equation}\label{vg}
V(|\phi|)={\frac{e^2}{2}}{\frac{(v^2-|\phi|^2)^2}{G(|\phi|)}}\,,
\end{equation}
and supposes that the fields satisfy
\bes \label{bps}
\begin{eqnarray}
D_{1}{\phi}\pm iD_{2}{\phi=0}\,,\label{bps1}\\
B \mp {\frac{e(v^2-|\phi|^2)}{G(|\phi|)}}=0 \label{bps2}\,,
\end{eqnarray}
\ees
one can write the energy of the field configurations as
\begin{equation}\label{energybo}
E_B=2{\pi}v^2|n|\,.
\end{equation}
The procedure leads to the first-order Eqs.~\eqref{bps} and energy \eqref{energybo} for the specific potential \eqref{vg}, so we conclude that the generalized model admits a first-order formalism if the magnetic permeability $G$ and the potential $V$ are related via the constraint \eqref{vg}.

In order to prepare the model numerical investigation, from now on we consider $e=v=1$. Moreover, with the ansatz given by Eqs.~\eqref{anz}, the above first-order Eqs.~\eqref{bps} become
\bes\label{25}
\begin{eqnarray}
g' &=& \pm \frac{ag}{r},\label{25a} \\
\frac{a'}{r} &=& \mp \frac{1-|\phi|^2}{G(|\phi|)}. \label{25b}
\end{eqnarray}
\ees
Also, we can write the energy density in the form
\be \label{dener11}
{\large\varepsilon}=2V(|\phi|)+\frac{2a^2g^2}{r^2},\,
\ee
where $V(|\phi|)$ has to obey Eq.~\eqref{vg}, now with $e=v=1$.

%%%%%%%%%%%%%%%%%%%%%%%%%%%%%%%%%%%%%%%%%%%%
\subsection{Standard vortices}\label{sec3}

The standard Maxwell-Higgs model is obtained in the limit $G(|\phi|) \to 1$. In this case, the Eqs. \eqref{eqm} become
\bes
\begin{eqnarray}
{\frac{1}{r}}[rg']'-\frac{a^2g}{r^2}-{\frac{1}{2}}V_{g} &=& 0\,,\\
r{\left[\frac{a'}{r}\right]}'-2g^2a &=& 0\,,
\end{eqnarray}
\ees
and reproduce the equations of motion of the standard model. According to our conventions, here we are dealing with scalar and vector fields with the same mass, and we can write the first order equations as
\begin{eqnarray}
g' = \pm \frac{ag}{r},\ \ \ \ \ \
\frac{a'}{r} = \mp (1-g^2). 
\end{eqnarray}

The solutions of the above equations have energy minimized to the Bogomol'nyi bound Eq.~\eqref{energybo}. This is well-known and can be found, for instance, in \cite{Vilenkin}. Below we will present numerical solutions to the above first-order equations to compare them with the vortices that appear in the new models that we now describe.

%%%%%%%%%%%%%%%%%%%%%%%%%%%%%%%%%%%%%%%%%%%%
\section{Generalized  Vortices}\label{sec3}
Let us now investigate some new models and their respective vortex solutions. We first suggest the function $G(|\phi|)$ and then write the corresponding potential, in order to study the vortex solutions, energy density and magnetic field.

%%%%%%%%%%%%%%%%%%%%%%%%%%%%%%%%%%%%%%%%%%%%
\subsection{A new model}

\begin{figure}[t]
\includegraphics[width=6cm,height=5cm]{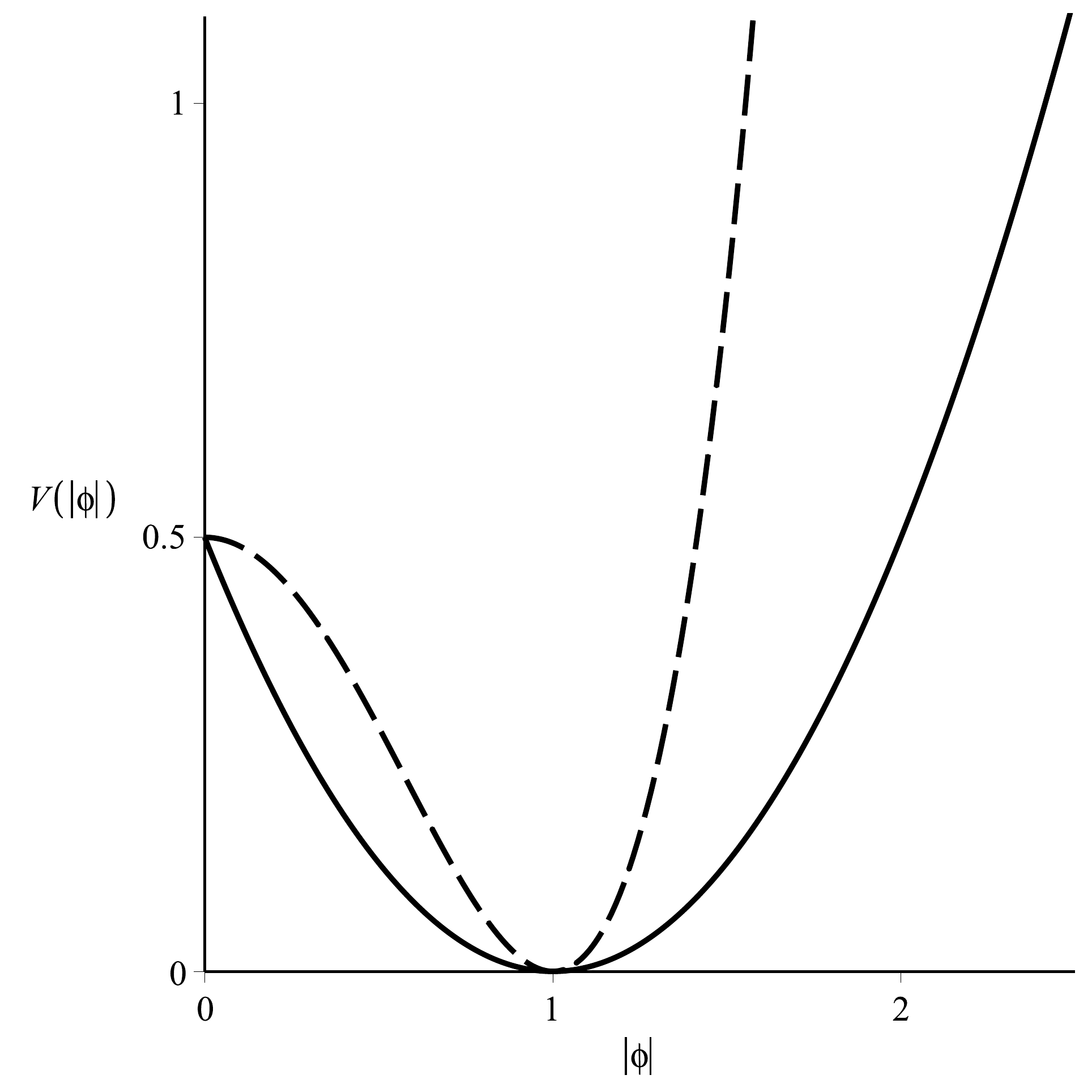} 
\caption{\footnotesize The potential \eqref{V1} and the standard Higgs potential, displayed with solid and dashed lines, respectively.}\label{fig1}
\end{figure}

Here we define $G(|\phi|)$ in the form
\begin{equation}\label{G1}
G(|\phi|)=\frac{(1-|\phi|^2)^2}{(1-|\phi|)^2}\,.
\end{equation}
We use the constraint give by Eq.~\eqref{vg} to get
\begin{equation}\label{V1}
V(|\phi|)=\frac{1}{2}(1-|\phi|)^2.
\end{equation}
This potential presents minima at $|\phi|=1$, and in Fig.~\ref{fig1} it is displayed together with the potential of standard Maxwell-Higgs model, for comparison.

%%%%%%%%%%%%%%%%%%%%%%%%%%%%%%%%%%%%%%%%%%%%%%%%%%%
%%%%%%%%%%%%%%%%%%%%%%%%%%%%%%%%%%%%%%%%%%%%%%%%%%%

The energy density {$\varepsilon$} and the magnetic field $B$ are given by
\be
{\Large\varepsilon}= (1-g)^2+\frac{2a^2g^2}{r^2}, \label{densy1}
\ee
and
\be 
B=\pm\frac{1-g}{1+g}. \label{B1}
\ee 
To see how they vary along the radial direction one has to solve the first-order equations 
\ben\label{50}
g' = \pm\frac{ag}{r},\;\;\;\;\;\frac{a'}{r} = \mp\frac{1-g}{1+g}.
\een

%%%%%%%%%%%%%%%%%%%%%%%%%%%%%
\begin{figure}[t]
{\includegraphics[width=4.2cm,height=4cm]{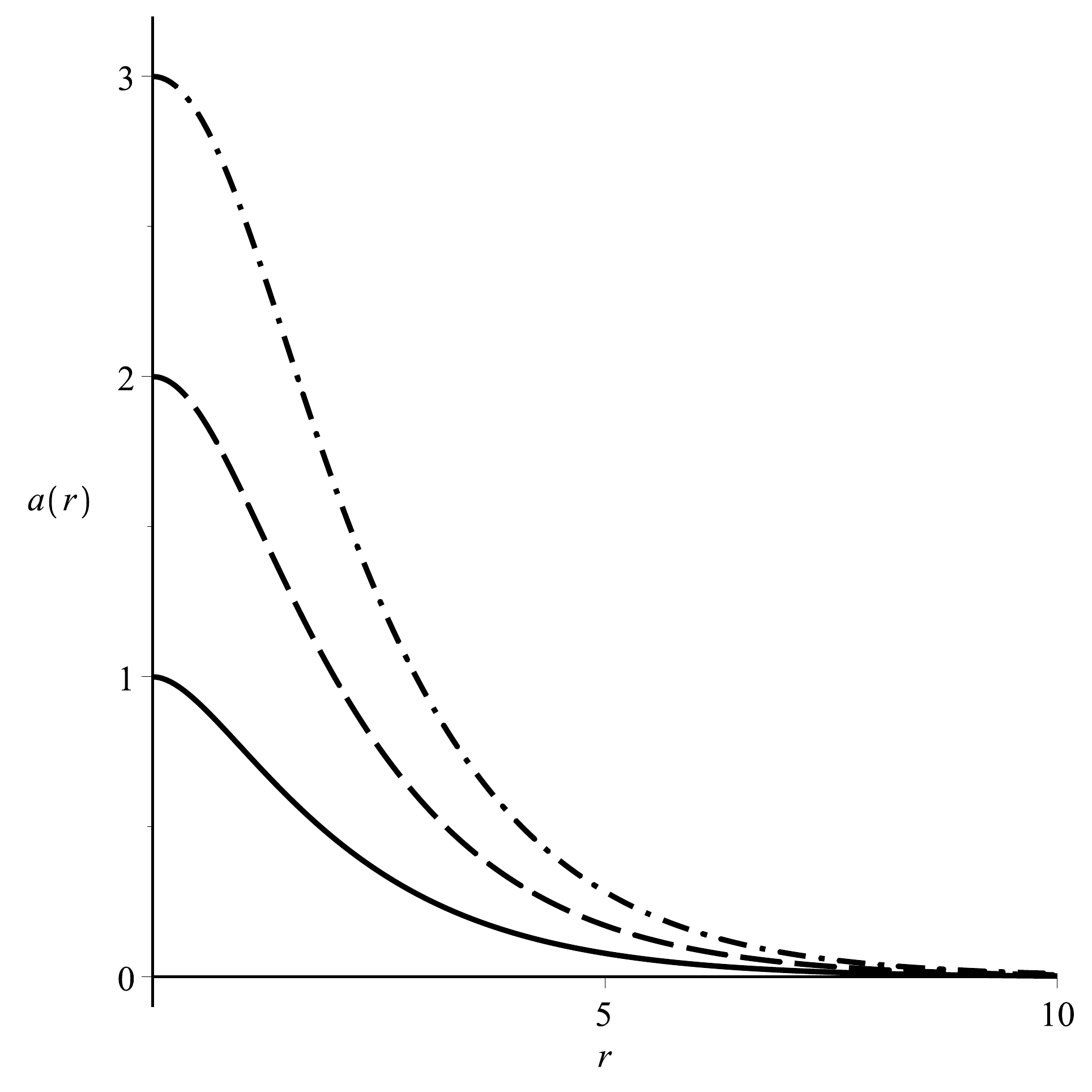}} 
{\includegraphics[width=4.2cm,height=4cm]{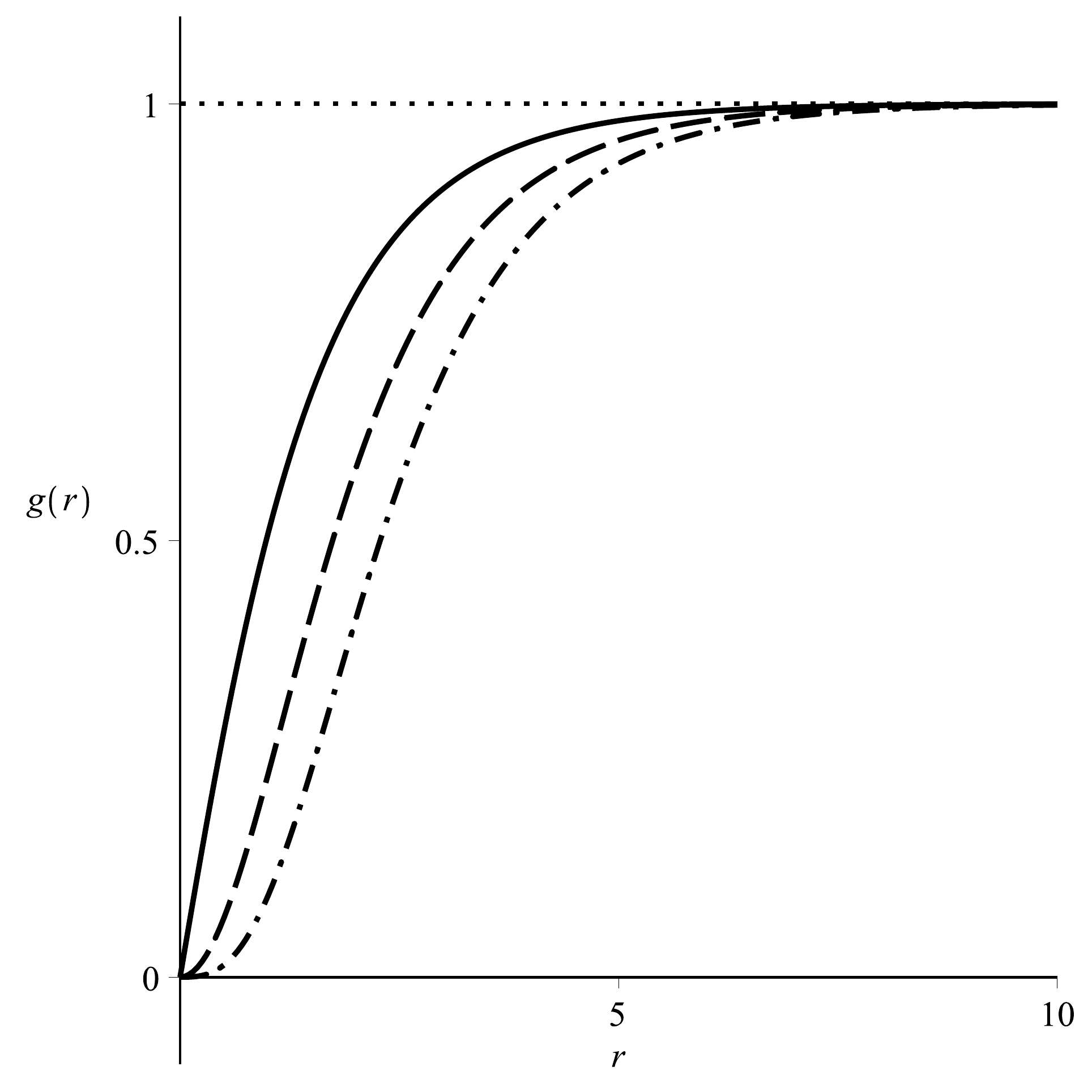}} 
{\includegraphics[width=4.2cm,height=4cm]{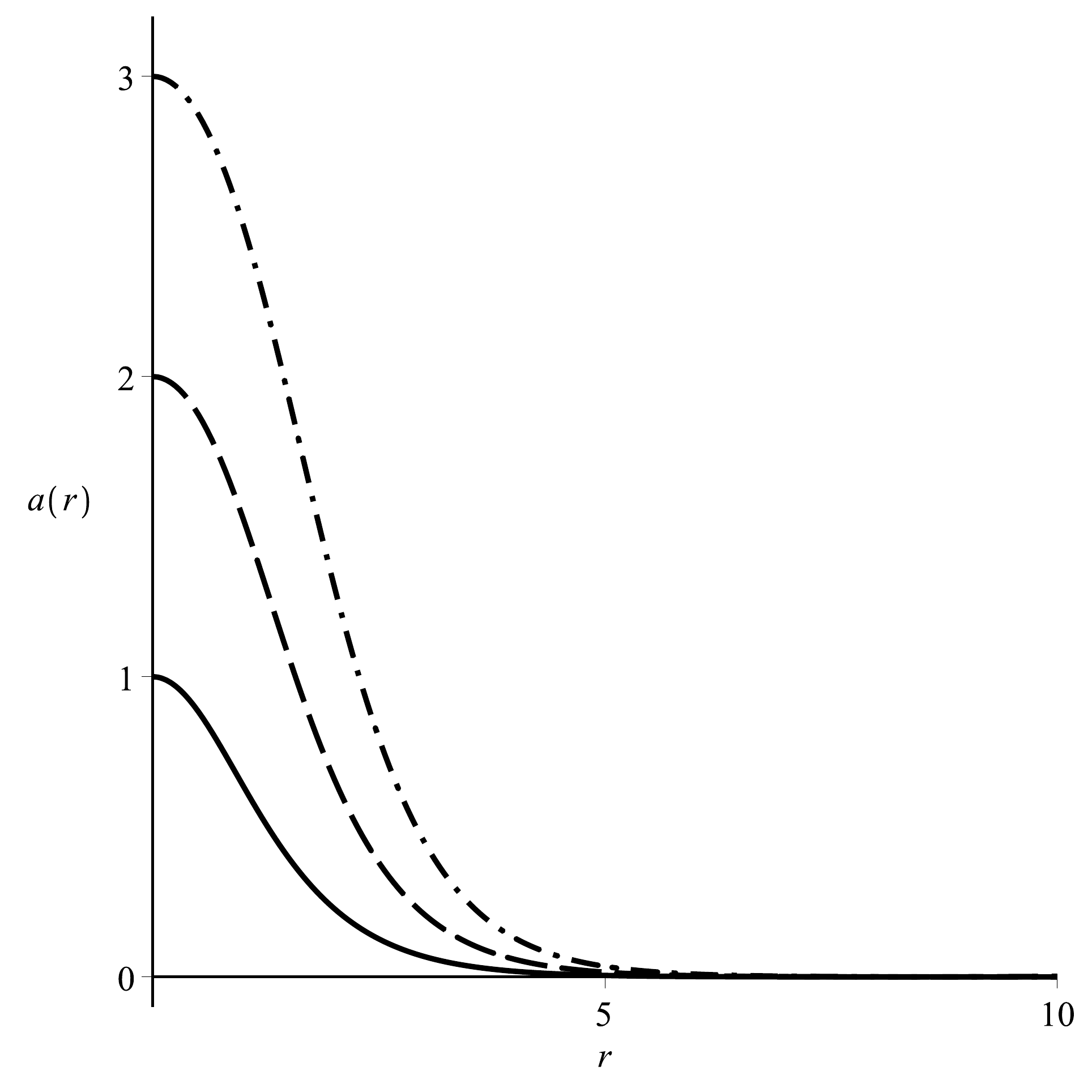}} 
{\includegraphics[width=4.2cm,height=4cm]{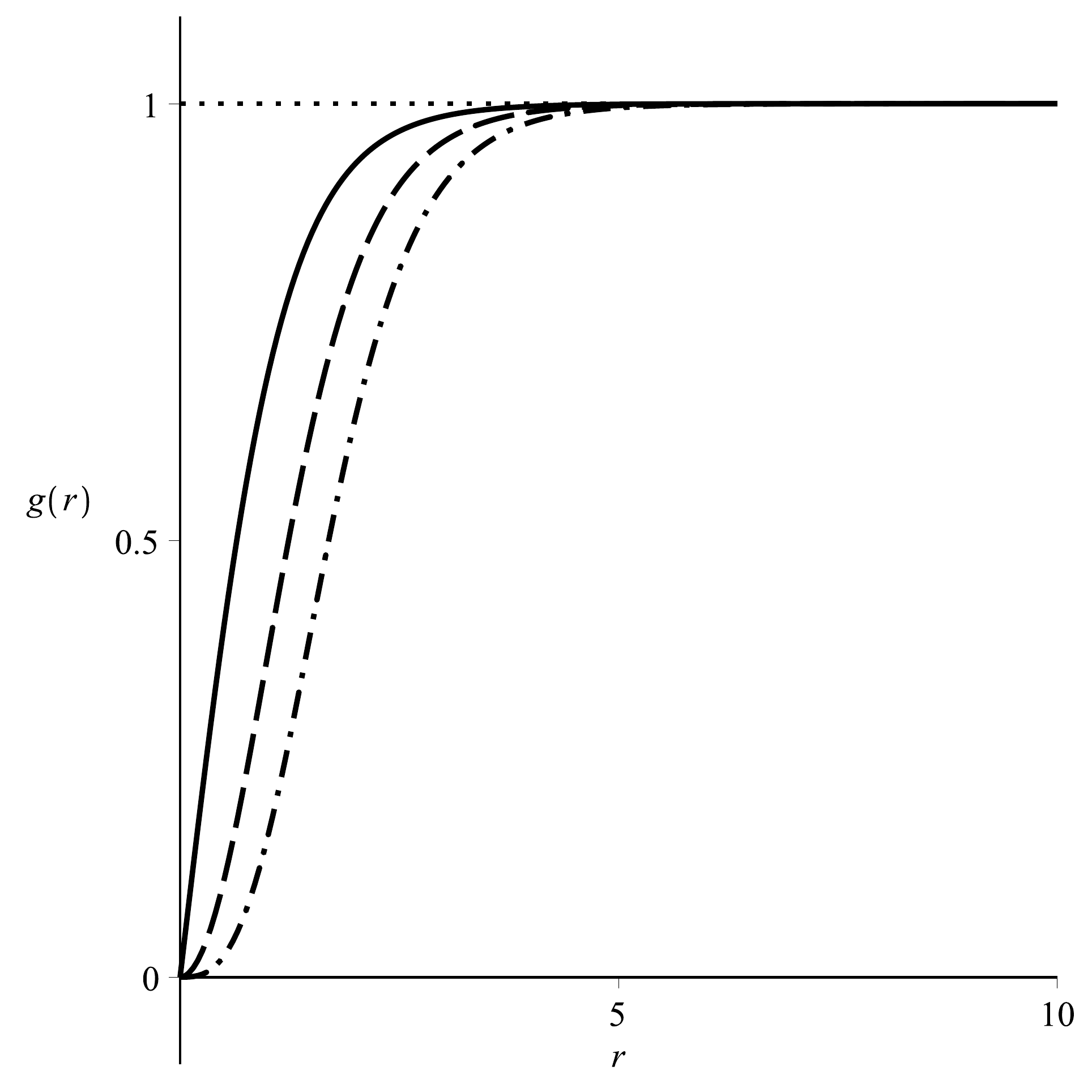}}
\caption{The functions $a(r)$ (left) and $g(r)$ (right), displayed for the new (top) and the standard Maxwell-Higgs (bottom) models. In all cases, the solid, dashed, and dot-dashed lines correspond to $n=1, 2,$ and $3$, respectively.} \label{fig2} 
\end{figure}
%%%%%%%%%%%%%%%%%%%%%%%%%%%%%%%

%%%%%%%%%%%%%%%%%%%%%%%%%%%%%%%%%%%%%%%%%
\begin{figure}[t]
{\includegraphics[width=4.2cm,height=4cm]{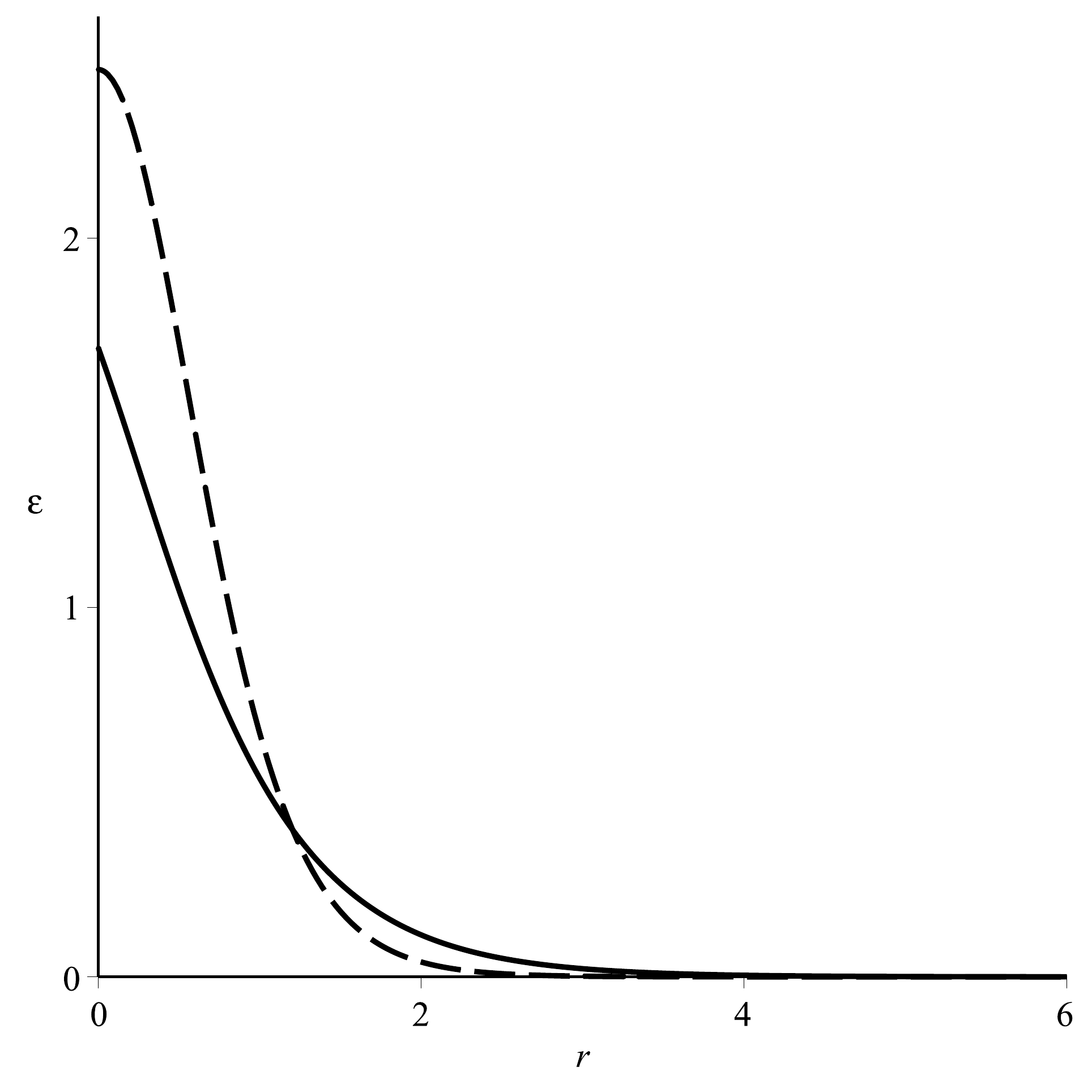}}
{\includegraphics[width=4.2cm,height=4cm]{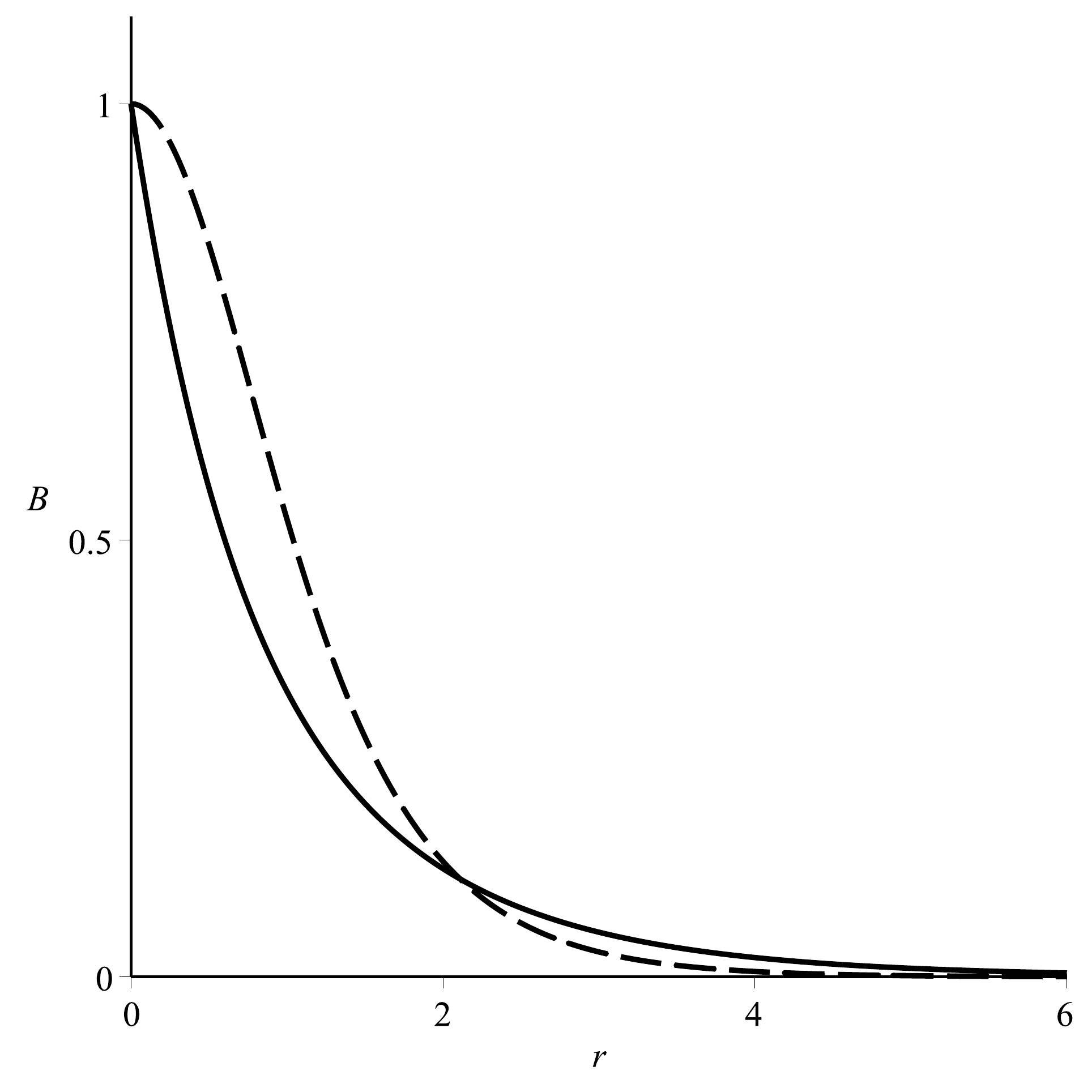}} 
\caption{The energy density (left) and the magnetic field (right) of the new and standard models, depicted for $n=1$ with solid and dashed lines, respectively.}\label{fig3}
\end{figure}
%%%%%%%%%%%%%%%%%%%%%%%%%%%%%%%%%%%%%%%%%%%%%%%%%%%

We have been unable to find analytical solution for the above equations, so we proceed with a numerical investigation. We first examine the asymptotic behavior of the solutions. We know that at larger distances, in the limit $r \to \infty$, the Eqs.~\eqref{50} can be approximated by the equations
$\delta'=\mp a/r$ and $a'=\mp \delta/2$, where $\delta=1-g$ is a very small quantity. The solutions with appropriate behavior at infinity are
\bes\label{agasym}
\begin{eqnarray}\label{gasym}
g&=&1-CK_{0}(r/{\scriptstyle \sqrt{2}})\,,\\
a&=&CrK_{1}({r/{\scriptstyle\sqrt{2}}})\,.\label{aasym}
\end{eqnarray}
\ees

To find the behavior of the solutions at small values of $r$, we attempt a power series solution and obtain, for positive $n$,
\bes
\begin{eqnarray}
g&=&Ar^n - \frac{Ar^{n+2}}{2(n+1)}+\scriptsize{{\cal{O}}(r^{2n+2})}\,,\\
a&=&n-\frac{r^2}{2}+\frac{2Ar^{n+2}}{n+2}+ {\cal{O}}(r^{\scriptsize{2n+2}}), 
\end{eqnarray}
\ees
where the constant $A$ is to be determined numerically to match the behavior of the solutions for larger values of $r$. To solve Eqs.~\eqref{50} numerically, we choose an initial value of $A$ and then integrate, searching to get the appropriate behavior for very large values of $r$. We then repeat the procedure with a new value of $A$ until find the correct value for $A$ that meets the above conditions. 

The results for $a(r)$ and $g(r)$ are shown in Fig.~\ref{fig2}, where we also display the solutions of the Maxwell-Higgs model. We note that the solutions of the new model are larger than they appear in the standard model. Moreover, in Fig.~\ref{fig3} one displays the energy density
{$\varepsilon$} and the magnetic field $B$ of both the new and the standard Maxwell-Higgs models, for comparison. And there one notes the same behavior, the energy density and the magnetic field of the new model seem to spread over a larger region in the plane.  
%%%%%%%%%%%%%%%%%%%%%%%%%%%%%%%%%%%%%%%%%%%%%%%%%%%

%%%%%%%%%%%%%%%%%%%%%%%%%%%%%%%%%%%%%%%%%%%%
\subsection{Another model}

We can choose another function $G(|\phi|)$ to describe the system. One considers the possibility
\begin{equation}\label{G2}
G(|\phi|)=\frac{(1-|\phi|^2)^2}{|\phi|^2(1-|\phi|)^2}.
\end{equation}
In this case, the constraint given by Eq.~\eqref{vg} leads us to the potential 
\begin{equation}\label{V2}
V(|\phi|)=\frac{1}{2}|\phi|^2(1-|\phi|)^2.
\end{equation}

\begin{figure}[t]
\includegraphics[width=6cm,height=5cm]{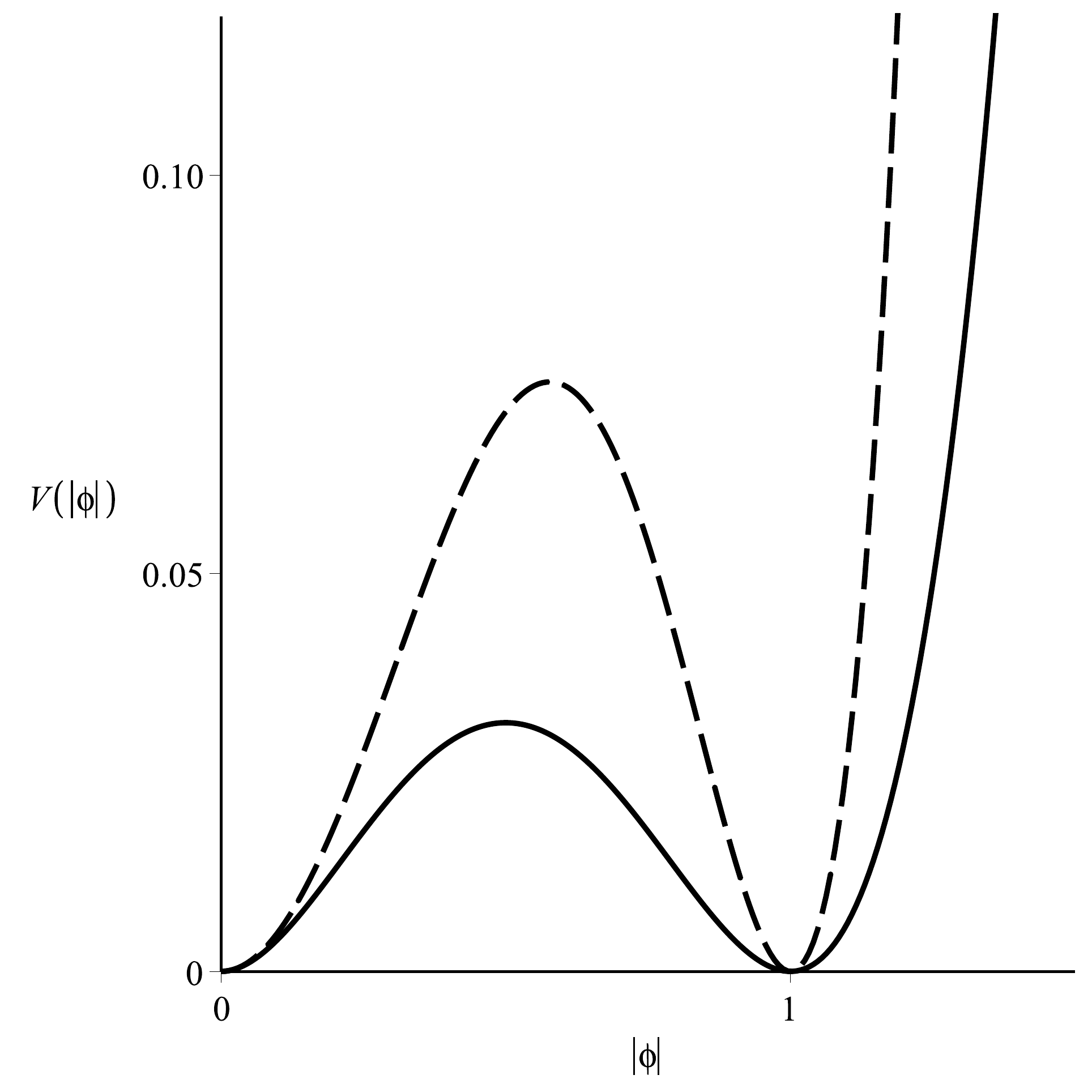} 
\caption{The two potentials \eqref{V2} and \eqref{v6}, displayed with solid and dashed lines, respectively.}
\label{fig4}
\end{figure}

This potential is of interest since it is of the forth-order power in the scalar field, but it resembles the sixth-order power potential that appears in the Chern-Simons model \cite{4,5}, which has the form
\be\label{v6}
V(|\phi|)=\frac{1}{2}|\phi|^2(1-|\phi|^2)^2.
\ee
They both have asymmetric minima at $|\phi|=1$ and one symmetric minimum at $|\phi|=0$, as one illustrates in Fig.~\ref{fig4}.
 
In this new model, we can write the energy density {$\varepsilon$} and the magnetic field $B$ in the form
\be
{\varepsilon}= g^2(1-g)^2+\frac{2a^2g^2}{r^2},
\ee
and
\be
B=\frac{g^2(1-g)}{1+g}\,.\label{B2}
\ee 
Moreover, the first-order equations are now given by
\be
g' = \pm\frac{ag}{r},\;\;\;\;\;\frac{a'}{r} = \mp\frac{g^2(1-g)}{1+g}.
\ee
To find the solution we proceed as before: we first investigate the asymptotic behavior, noting that the fields go as they did in the previous model, as described by the Eqs.~\eqref{agasym} in the limit $r \to \infty$. However, near the origin the power series are 
\bes
\begin{eqnarray}
g&=&Ar^n - \frac{A^2r^{3n+2}}{2(n+1)(3n+2)}+ {\cal{O}}(r^{4n+2})\,,\\
a&=&n-\frac{A^2r^{2n+2}}{2n+2}+\frac{2A^3r^{3n+2}}{3n+2} + {\cal{O}}(r^{4n+2})\,.
\end{eqnarray}
\ees

%%%%%%%%%%%%%%%%%%%%%%%%%%%%%%%%%%%%%%%%%%%%%%%%%%
%%%%%%%%%%%%%%%%%%%%%%%%%%%%%%%%%%%%%%%%%%%%%%%%%%%

\begin{figure}[t]
{\includegraphics[width=4.2cm,height=4cm]{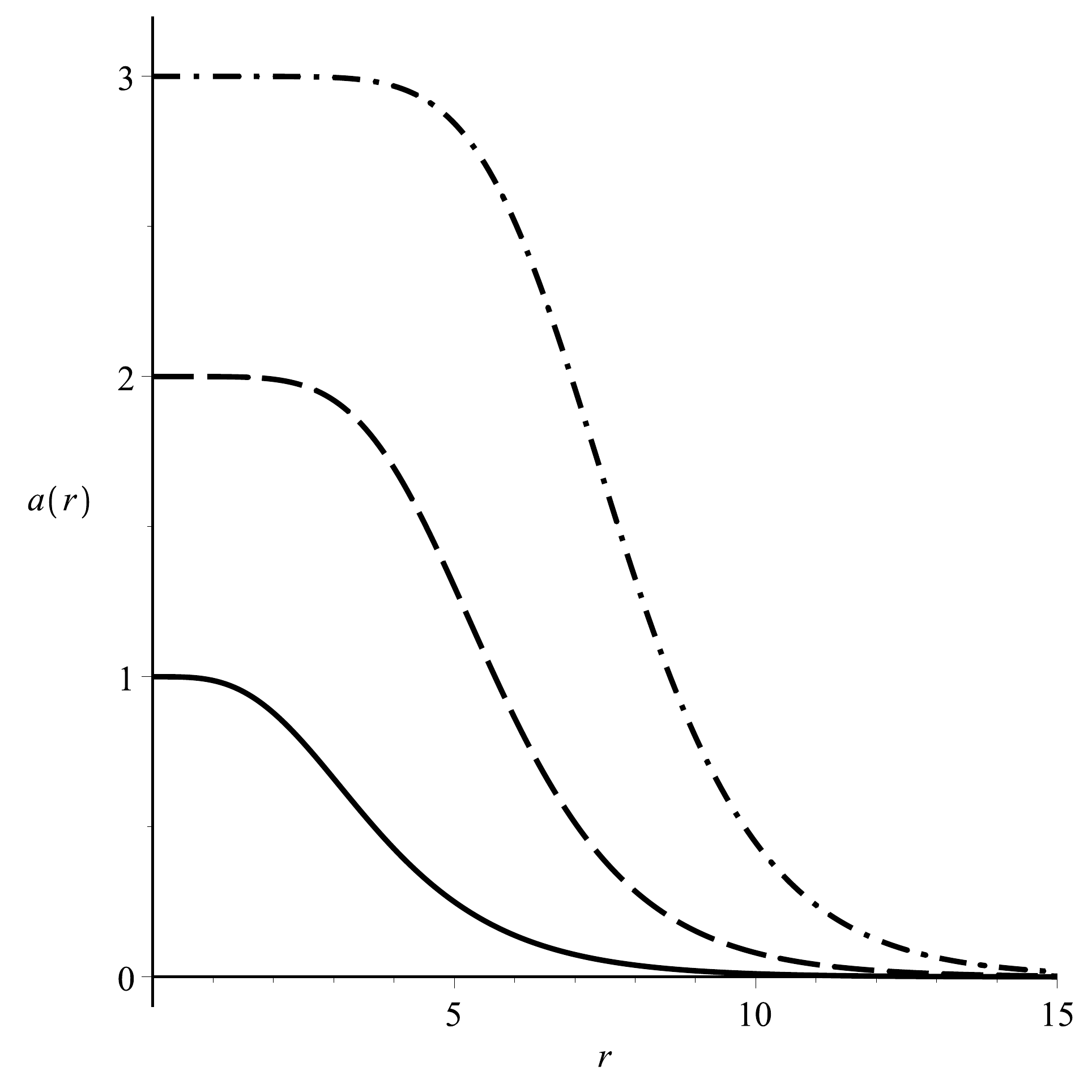}} 
{\includegraphics[width=4.2cm,height=4cm]{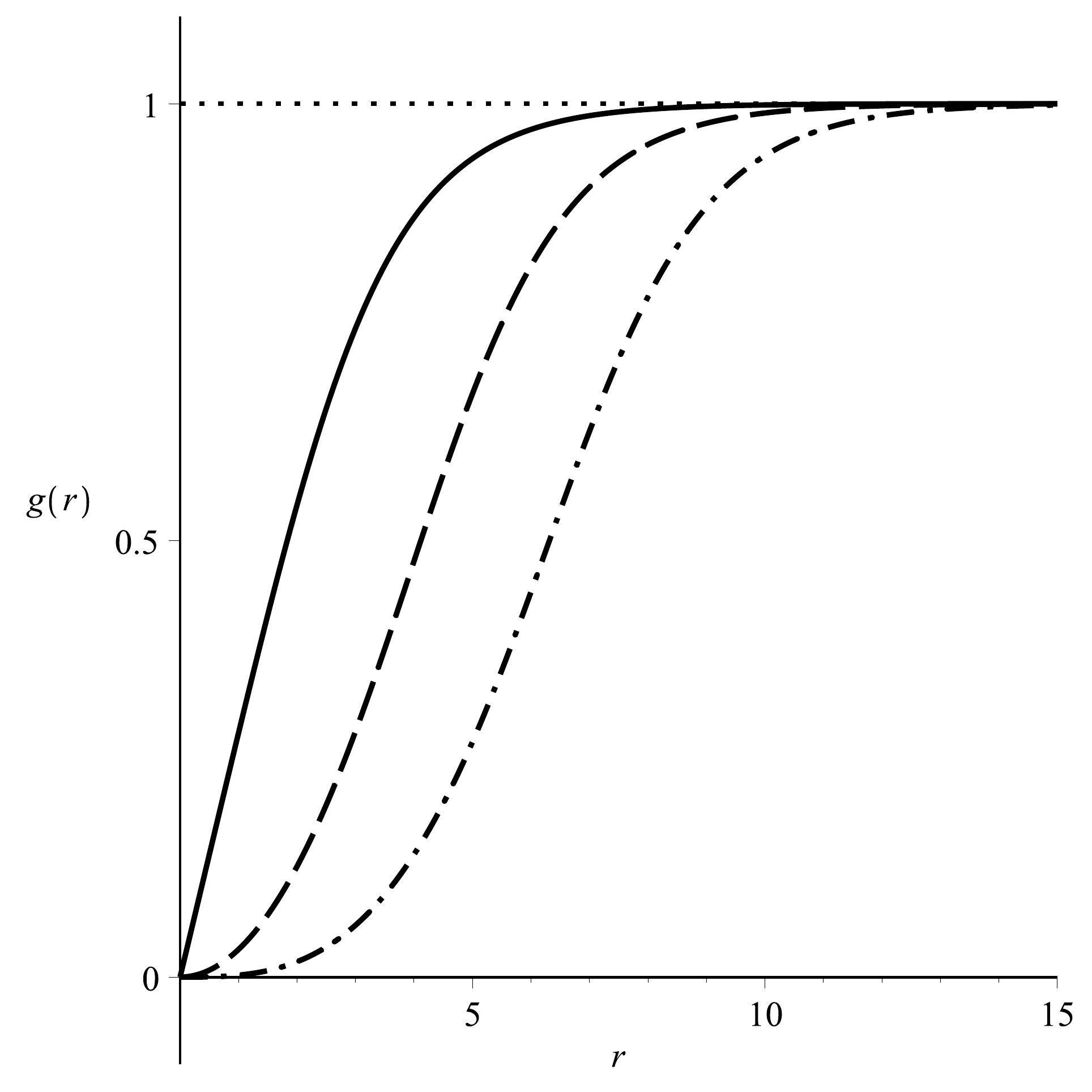}} 
{\includegraphics[width=4.2cm,height=4cm]{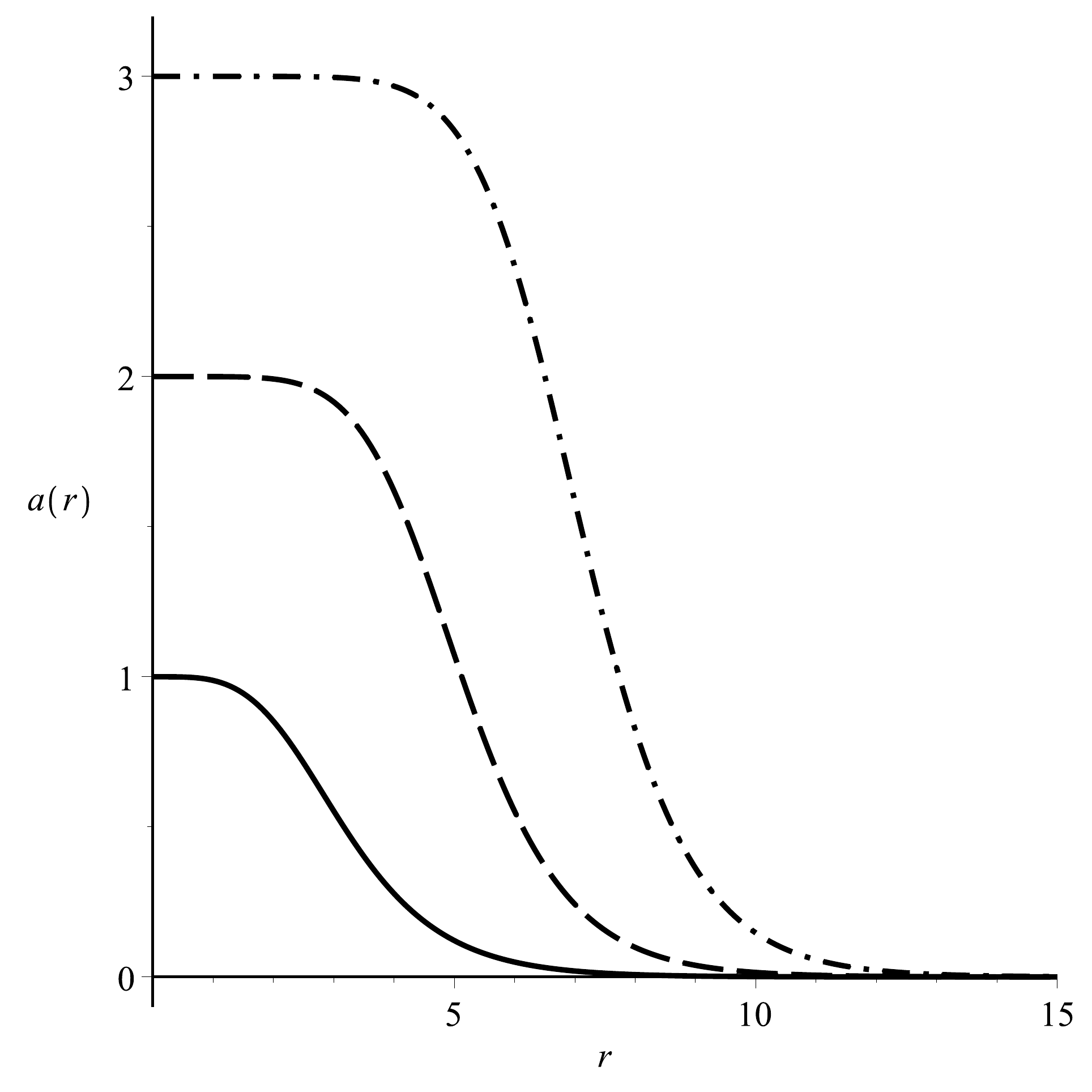}} 
{\includegraphics[width=4.2cm,height=4cm]{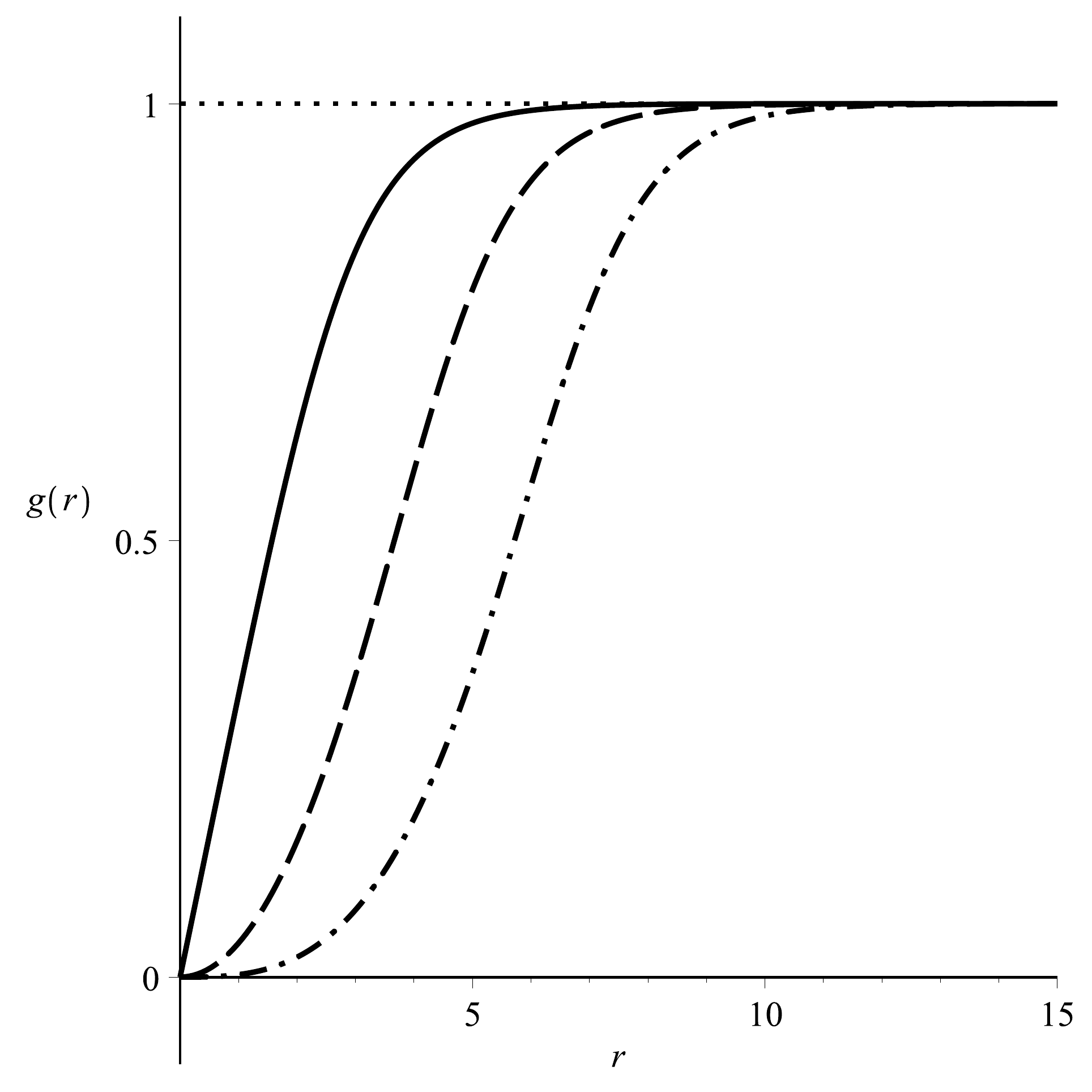}}
\caption{The functions $a(r)$ (left) and $g(r)$ (right) for the model \eqref{V2} (top) and for the Chern-Simons model \eqref{v6} (bottom). The solid, dashed, and dot-dashed lines correspond to $ n=1,2,3 $, respectively.} \label{fig5} 
\end{figure}

\begin{figure}[t]
{\includegraphics[width=4.2cm,height=4cm]{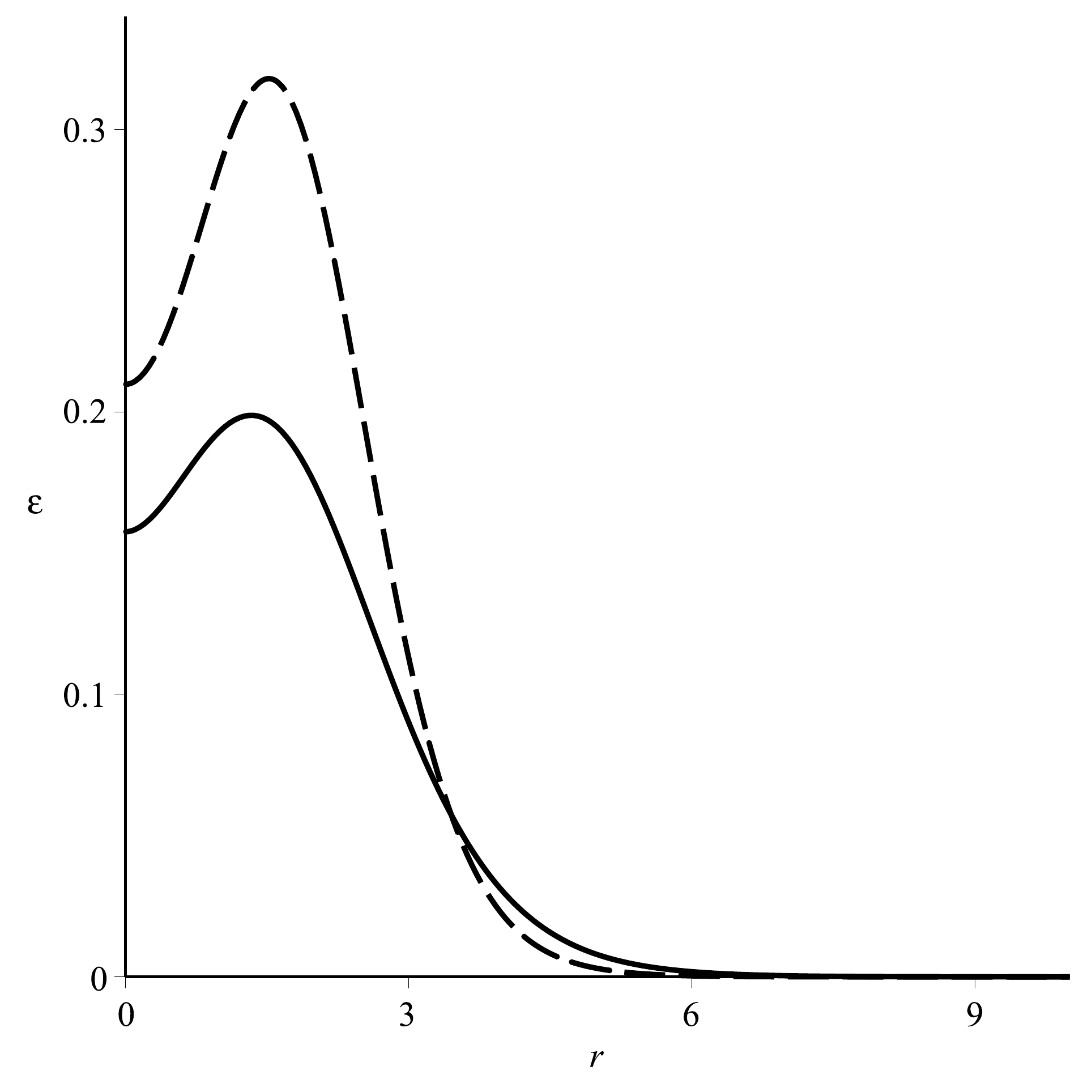}} 
{\includegraphics[width=4.2cm,height=4cm]{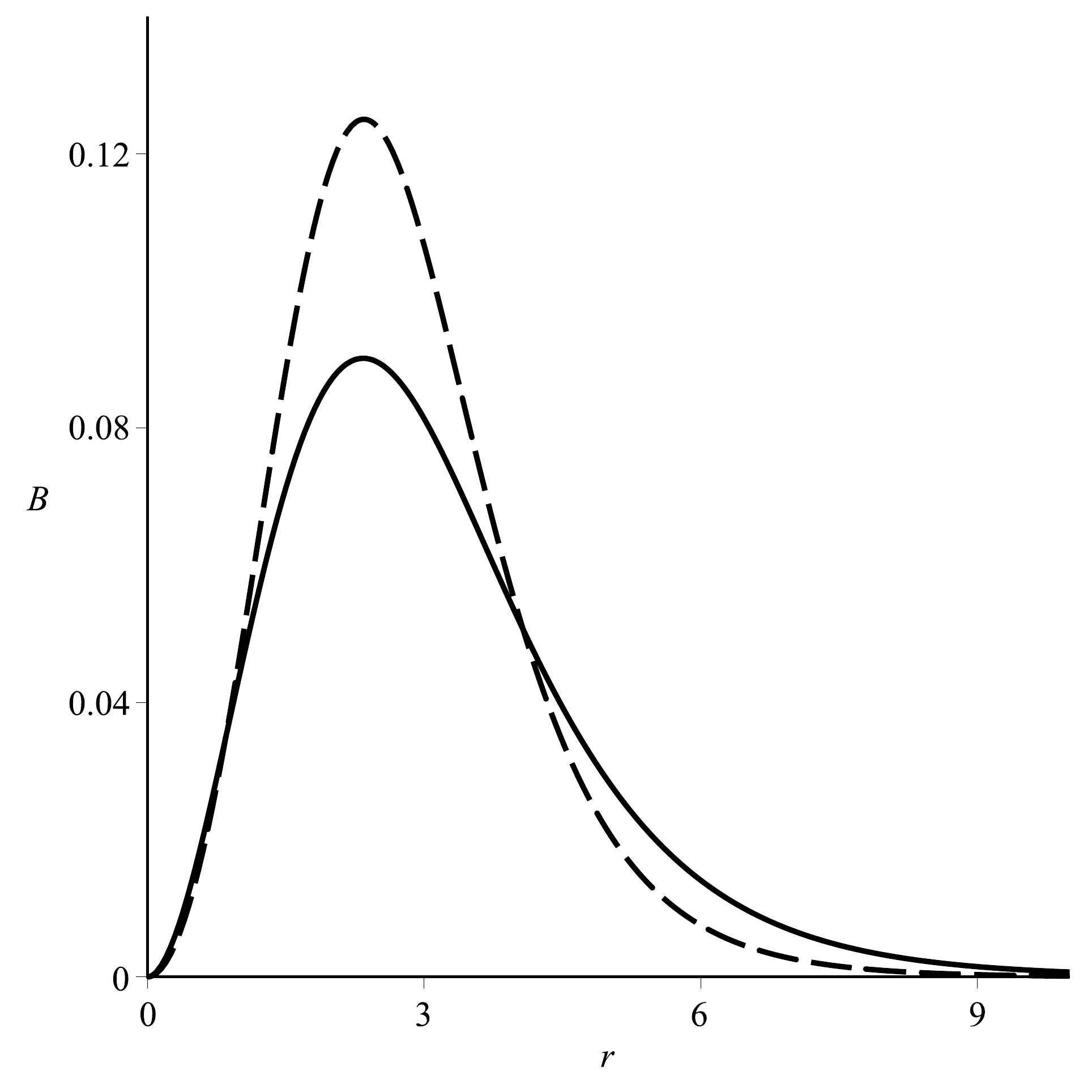}} 
\caption{The energy density (left) and the magnetic field (right) for $n=1$, for the model \eqref{V2} and for the Chern-Simons model, depicted with solid and dashed lines, respectively.} \label{fig6}
\end{figure}

We use these results to integrate numerically the first-order equations. The behavior of the fields are then displayed in Fig.~\ref{fig5}, where we also show the results for the Chern-Simons model, for comparison. Moreover, in Fig.~\ref{fig6} we display the behavior of energy density {$\varepsilon$} and the magnetic field $B$ for the model \eqref{V2} and for the Chern-Simons model. As in the previous model, one notes here that the solutions, energy density and magnetic field of the new model also spread over a larger region in the plane, if compared with the Chern-Simons case.

%%%%%%%%%%%%%%%%%%%%%%%%%%%%%%%%%%%%%%%%%%%%%%%%%%%
%%%%%%%%%%%%%%%%%%%%%%%%%%%%%%%%%%%%%%%%%%%%

%%%%%%%%%%%%%%%%%%%%%%%%%%%%%%%%%%
%%%%%%%%%%%%%%%%%%%%%%%%%%%%%%%%%%%%%%%%%%%%%%%%%%%
\section{Compact Vortices}\label{sec4}

We now turn attention to the possibility of constructing compact vortices. One first notes from the results of the previous Sec.~\ref{sec3} that it is possible to modify the function $G(\phi)$ in order to change the potential of the model. Thus, we get inspiration from the recent work on compact kinks \cite{CK} to describe a route to build compact vortices. We recall that in \cite{CK} one developed the possibility of changing the scalar field self-interactions, in a way capable of shrinking the solution to a compact interval of the real line. We use the same idea here, and below we illustrate this possibility introducing two distinct models.

Before going on the subject, however, one searches to recall recent efforts to describe compact vortices. In Ref.~\cite{Babi} the author deals with the same issue, but there one considers a non-canonical kinetic term, leading to a different scenario. A similar investigation, with models also containing non-canonical kinetic terms has been carried out in \cite{oli}. However, one notes that both the energy density and the magnetic field do not respond as significantly as the solutions do. 

These results motivate us to revisit the subject, with focus on the construction of generalized models that support genuine compact vortices, with the energy density and magnetic field vanishing outside a compact interval of the radial coordinate. We implement this possibility below, investigating two distinct models that support compact vortices.

%%%%%%%%%%%%%%%%%%%%%%%%%%%%% 
\subsection{A model for compact vortices}

\begin{figure}[t]
{\includegraphics[width=5.6cm]{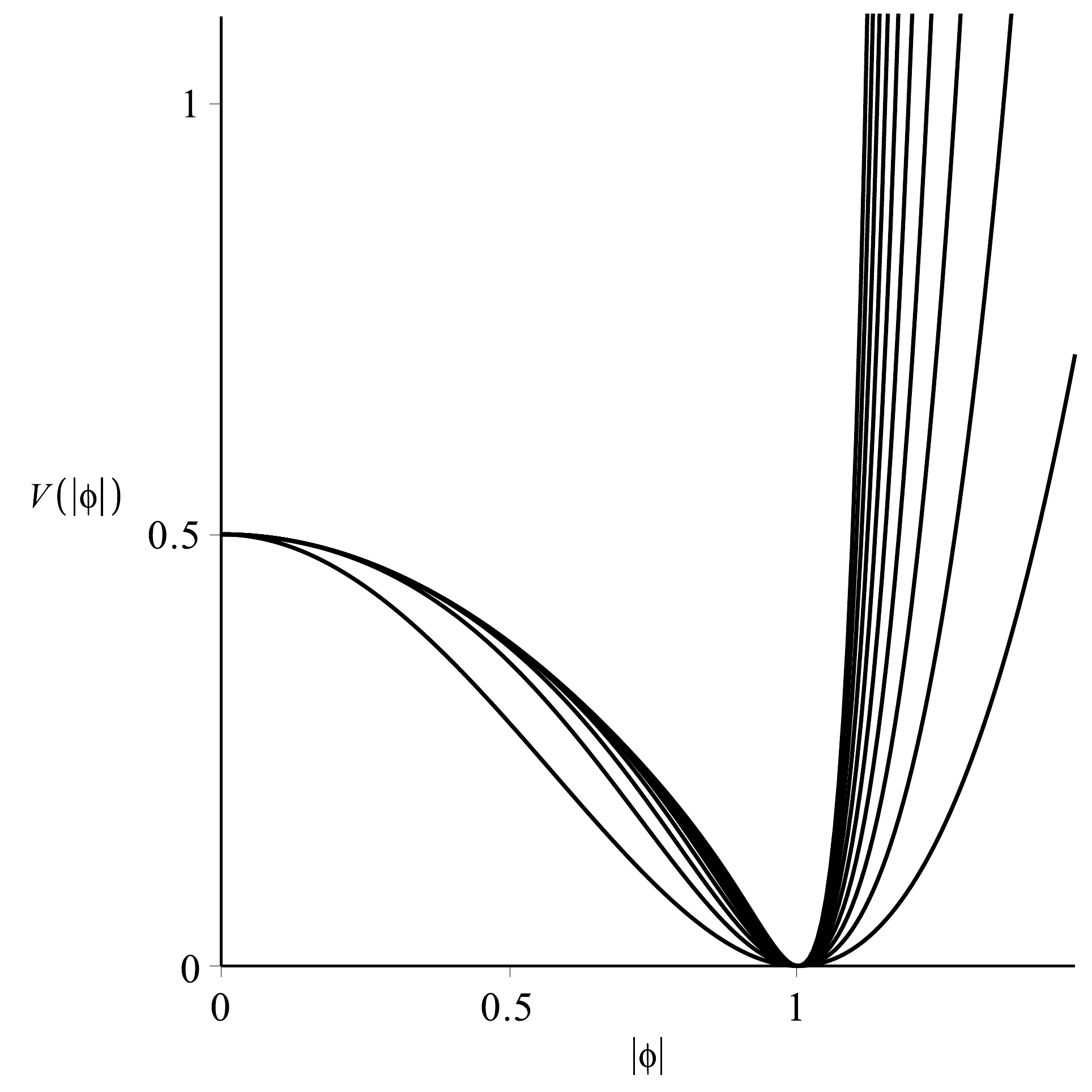}} 
\caption{The potential \eqref{Vc} for $l=1,2,3,\dots, 10$.} 
\label{fig7}
\end{figure}

We follow as in the previous section and choose the magnetic permeability in the form
\begin{equation}\label{Gc}
G(|\phi|)=\frac{1-|\phi|^2}{1-|\phi|^{2l}},
\end{equation}
where $l$ is a positive real parameter, such that $l\geq1$. With this choice, the constraint \eqref{vg} leads to the potential 
\begin{equation}\label{Vc}
V(|\phi|)=\frac12(1-|\phi|^2)(1-|\phi|^{2l}).
\end{equation}
Note that the case $l=1$ leads us back to the standard Maxwell-Higgs model. This new potential has a local maximum at $|\phi|=0$, with $V(0)=1/2$, and the minima are all located at $|\phi|=1$. This is similar to the standard model, but now the parameter $l$ introduces a nice behavior, as we show in Fig.~\ref{fig7}.

%%%%%%%%%%%%%%%%%%%%%%%%%%%%
%%%%%%%%%%%%%%%%%%%%%%%%%%

In this case, the first-order equations become
\begin{equation}\label{eqac}
g'= \pm\frac{ag}{r}, \quad \quad
\frac{a^\prime}{r}= \mp (1-g^{2l}). 
\end{equation}
We first study the limit $r\to0$, considering $a(r)\approx n+a_0(r)$ and $g(r)\approx g_0(r)$ and going up to first order in $a_0(r)$ and
$g_0(r)$. The procedure leads to
\ben
a_0(r) = -\frac{r^2}{2},\quad \quad 
g_0(r) = \alpha r^n,
\een
where $\alpha$ is an integration constant. A similar analysis can be done for the asymptotic behavior, in the limit $r\to\infty$. We take $a(r)\approx a_{asy}(r)$ and $g(r) \approx 1+g_{asy}(r)$ in Eqs.~\eqref{eqac} to get
\bes
\ben
a_{asy}(r) &=& \sqrt{2l}\beta r K_1\left(\sqrt{2l}r\right), \\
g_{asy}(r) &=& -\beta K_0\left(\sqrt{2l}r\right),
\een 
\ees
where $K_\nu(x)$ is the modified Bessel function of the second kind and $\beta$ is an integration constant. For a general $l$, we must solve Eqs.~\eqref{eqac} numerically, since it is very complicated to find analytical solutions for the problem. Nevertheless, for a general $n$ and very large $l$, it is possible to show that the model supports the compact solutions
\bes\label{solc}
\ben
a_c(r)&=&
\begin{cases}
n-\frac{1}{2}r^2,\,\,\,&r\leq \sqrt{2n},\\
0, \,\,\, & r>\sqrt{2n}.
\end{cases} \\
g_c(r)&=&
\begin{cases}
\left(\frac{r}{\sqrt{2n}}\right)^n\, e^{(2n-r^2)/4},\,\,&r\leq \sqrt{2n},\\
1, \,\, & r>\sqrt{2n}  .
\end{cases}
\een
\ees

One can wonder if the solutions \eqref{solc} are compatible with the equations of motion and energy density, since a discontinuity issue may appear at the point $r=\sqrt{2n}$. We have checked the compact profile, and noted that in the generalized model the factor $G(|\phi|)$ work to regularize the behavior, since it vanishes for $|\phi|=1$. Thus, the compact limit is regular and the BPS bound \eqref{energybo} still holds for these compact solutions.

%%%%%%%%%%%%%%%%%%%%%%%
\begin{figure}[t]
{\includegraphics[width=5cm]{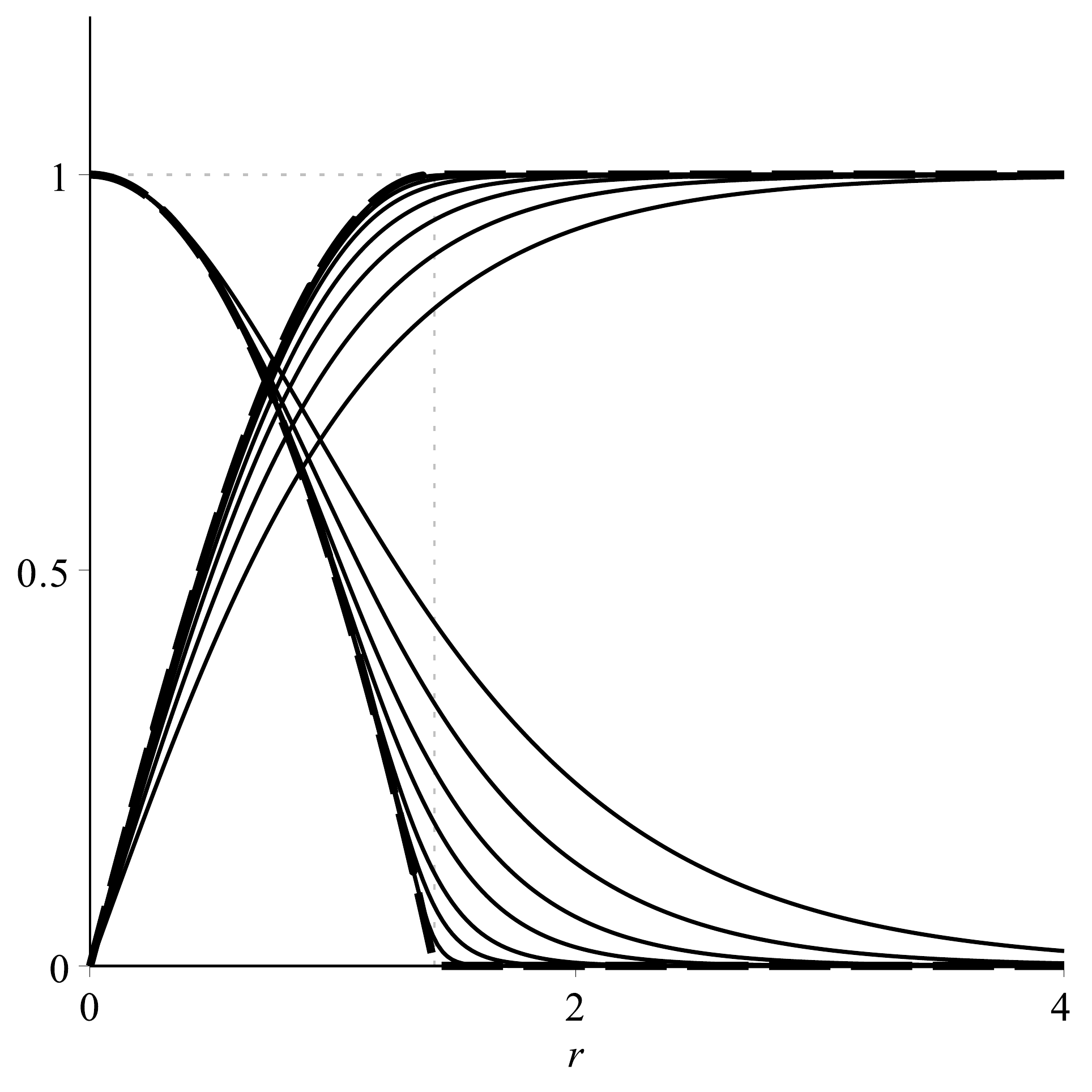}} 
\caption{The solutions $a(r)$ and $g(r)$ for $n=1$. We first consider $l=1$ and then increase it to larger and larger values. The dashed lines stand for the compact limit, given by Eqs.~\eqref{solc} with $n=1$.}\label{fig8}
\end{figure}
%%%%%%%%%%%%%%%%%%%%%%%%%

%%%%%%%%%%%%%%%%%%%%
\begin{figure}[t!]
{\includegraphics[width=4.2cm]{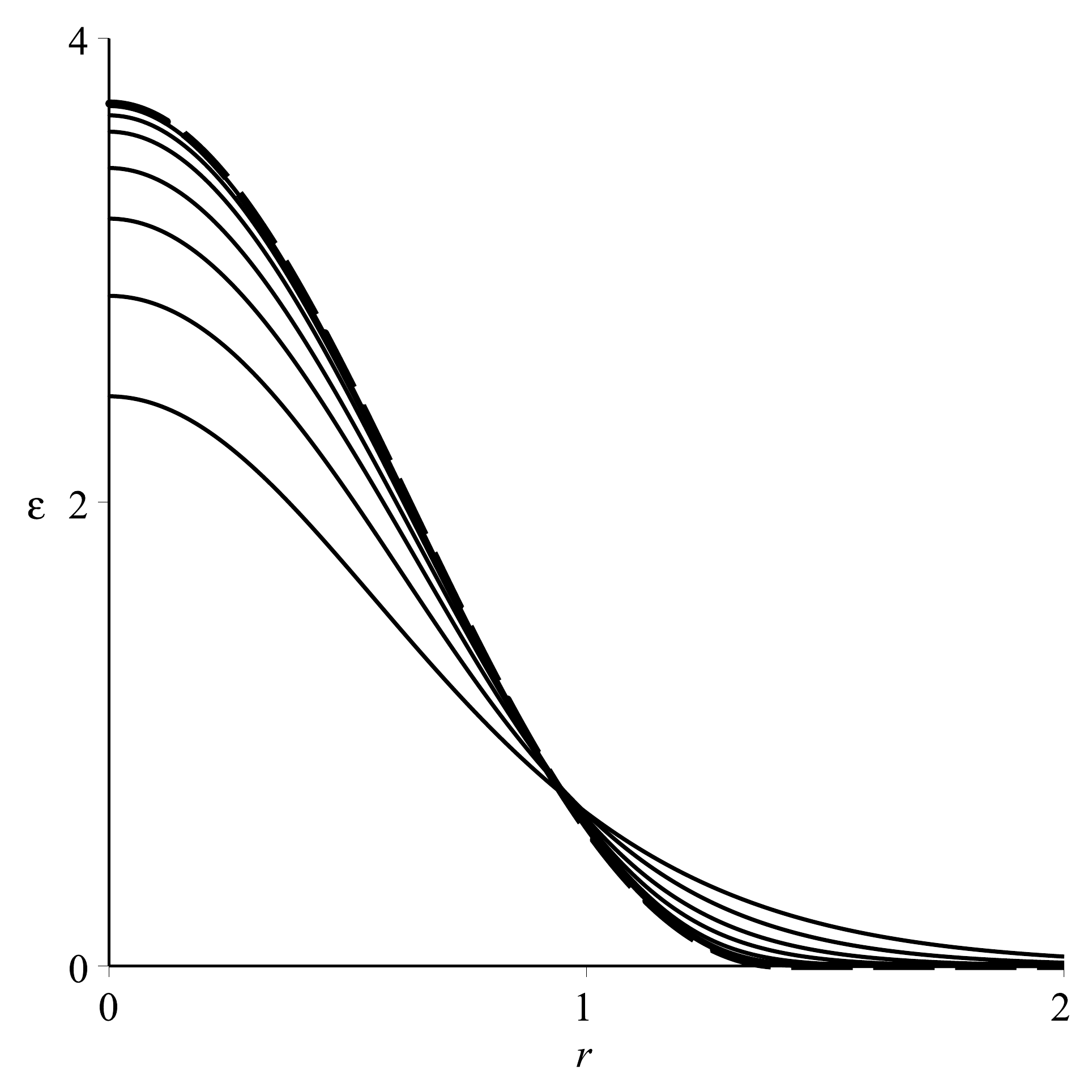}}
{\includegraphics[width=4.2cm]{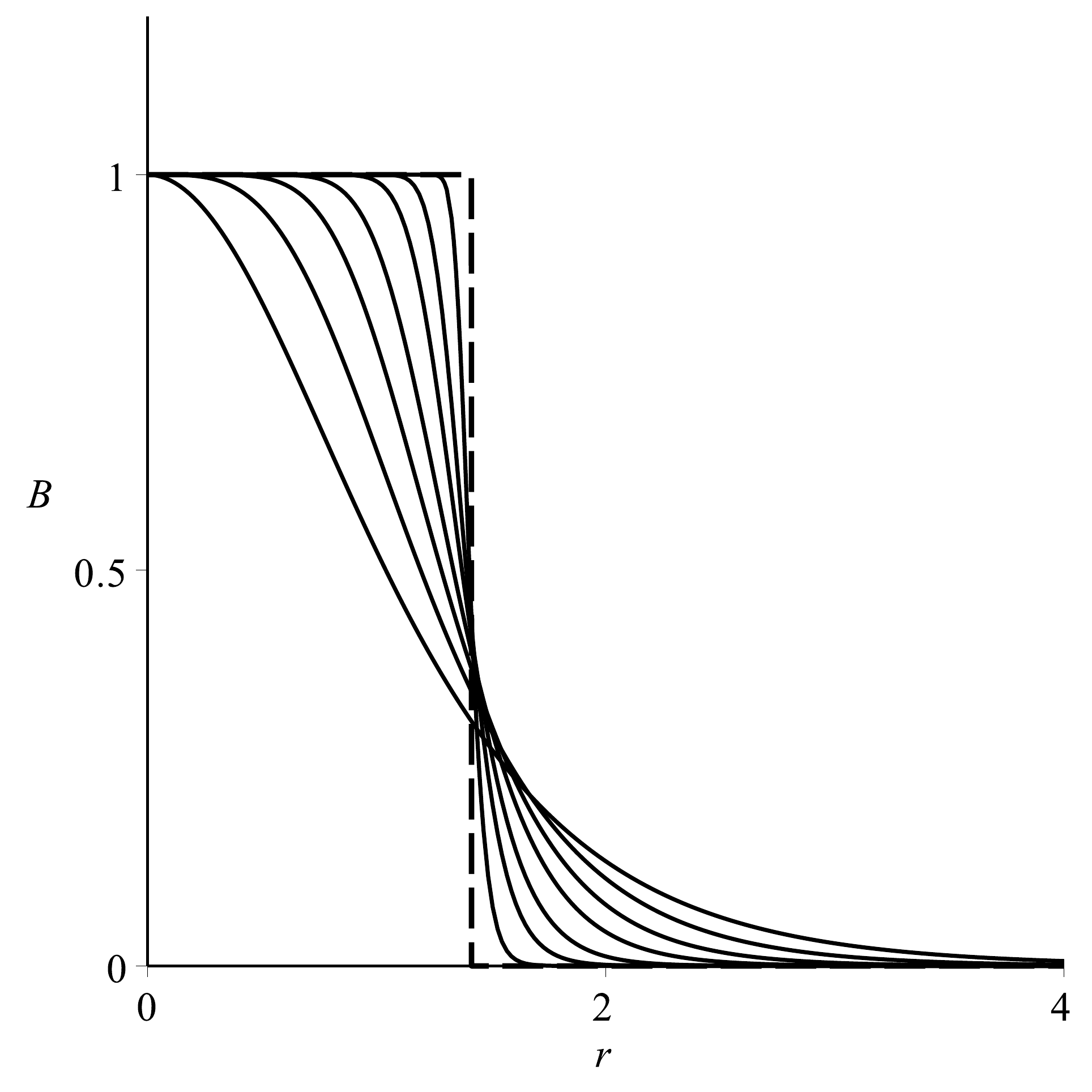}} 
\caption{The energy density (left), and the magnetic field (right) for $n=1$. We first consider $l=1$ and then increase it to larger and larger values. The dashed lines represent the compact limit, given by Eqs.~\eqref{dbc} with $n=1$.}\label{fig9}
\end{figure}
%%%%%%%%%%%%%%%%%%%%%%%

In Fig.~\ref{fig8}, we display the solutions for $n=1$ and for several values of $l$.
We have checked that the energy density and the magnetic field tend to become compact, and for $l$ very large one gets 
\bes\label{dbc}
\ben
\epsilon_c(r)&=&
\begin{cases}
{\scriptstyle{1- \left(2n+1 -\frac{r^2}{2} - \frac{2n^2}{r^2} \right)} \left(\frac{r^2}{2n}\right)^n e^{(2n-r^2)/2}},\,\, & {\scriptstyle{r\leq \sqrt{2n}}}\\
0, \,\, & {\scriptstyle{r> \sqrt{2n}}}
\end{cases} \nonumber \\
\\
B_c(r)&=&
\begin{cases}
1,\,\,\,&{\scriptstyle{r\leq \sqrt{2n}}}\\
0, \,\,\, & {\scriptstyle{r>\sqrt{2n}}}.
\end{cases}
\een
\ees

In Ref.~\cite{oli}, in particular, the route there proposed to shrink the vortex solutions to a compact interval was not able to make the energy density vanish outside the compact interval. This was perhaps the reason to call the solutions {\it compactlike} vortices. Here, the vortex becomes a compact solution, since both the energy density and magnetic field shrink to the compact interval. We illustrate this fact in Fig.~\ref{fig9}, displaying the energy density and the magnetic field for $n=1$, for several values of $l$. It is interesting to see that in the compact limit the magnetic field is constant inside the compact interval, so it seems to map the magnetic field of an infinitely long solenoid. 
As we commented before, the discontinuity in the magnetic field in the generalized model does not modify the energy density, due to the presence of the generalized magnetic permeability.

%%%%%%%%%%%%%%%%%%%%%%%%%%%%%%%%%%%%%%%%%%%%
\begin{figure}[t]
{\includegraphics[width=5.6cm]{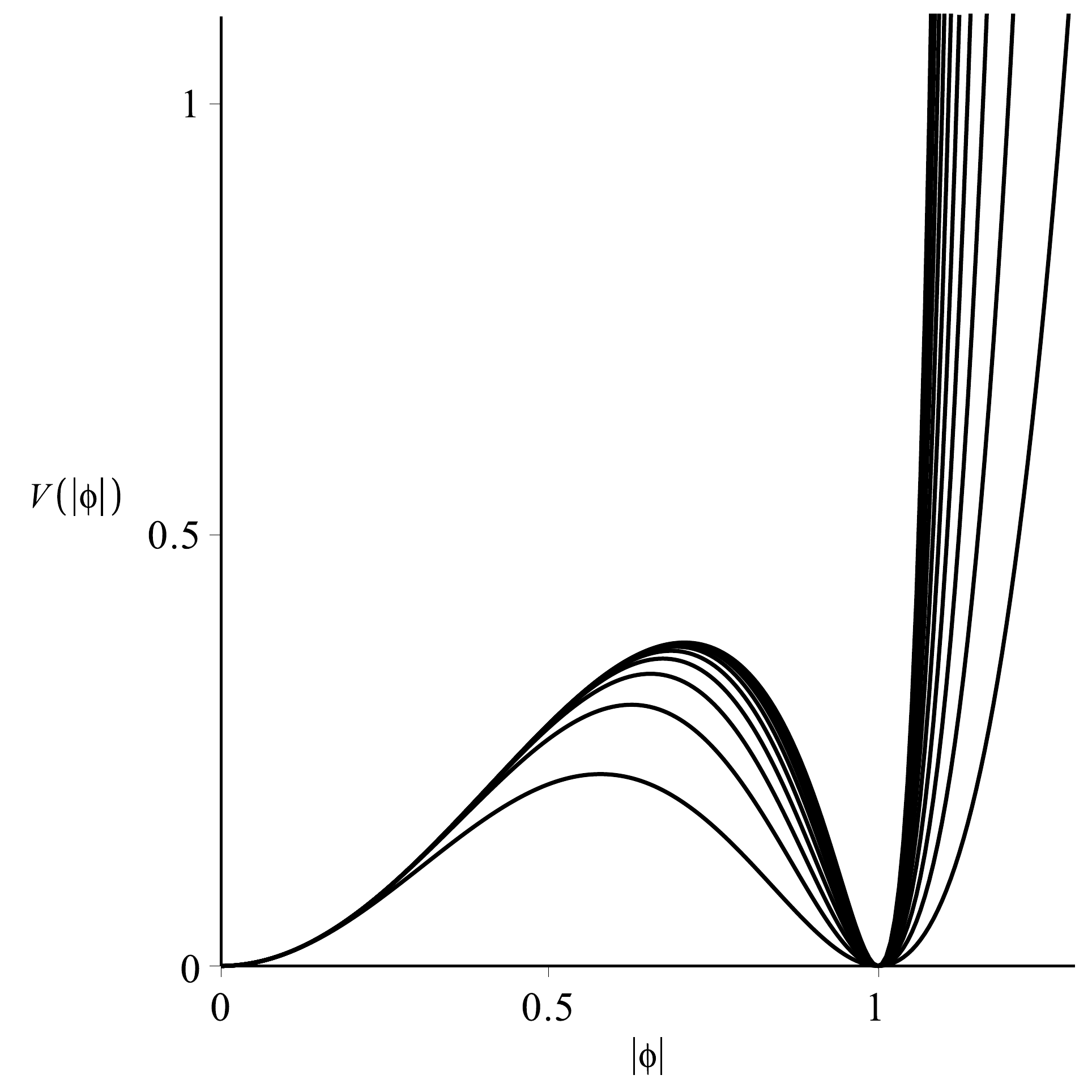}} 
\caption{The potential \eqref{Vhc} for $l=1,2,3,\dots, 10$.} 
\label{fig10}
\end{figure}
%%%%%%%%%%%%%%%%%%%%%%%%%%%%%%%%%%%%%%%%

%%%%%%%%%%%%%%%%%%%%%%%%%%%%%%%%%%%%%%%%%%%%%%%%% 
\subsection{Another model for compact vortices}

We can choose another model for the magnetic permeability. We consider the case
\begin{equation}\label{Ghc}
G(|\phi|)= \frac{1-|\phi|^2}{3|\phi|^2(1-|\phi|^{2l})},
\end{equation}
where $l$ is a positive real parameter, such that $l\geq1$. Then, the constraint given by \eqref{vg} implies that the potential has the form 
\begin{equation}\label{Vhc}
V(|\phi|)=\frac32 |\phi|^2(1-|\phi|^2)(1-|\phi|^{2l}).
\end{equation}
Note that the case $l=1$ gives the $|\phi|^6$ potential. This expression has a local minimum at $|\phi|=0$, and a set of maxima in between the minimum at zero and the set of minima at $|\phi|=1$; see Fig.~\ref{fig10}, where we display the potential \eqref{Vhc} for several values of $l$.

%%%%%%%%%%%%%%%%%%%%%%%%%%%%%%
\begin{figure}[t]
{\includegraphics[width=5cm]{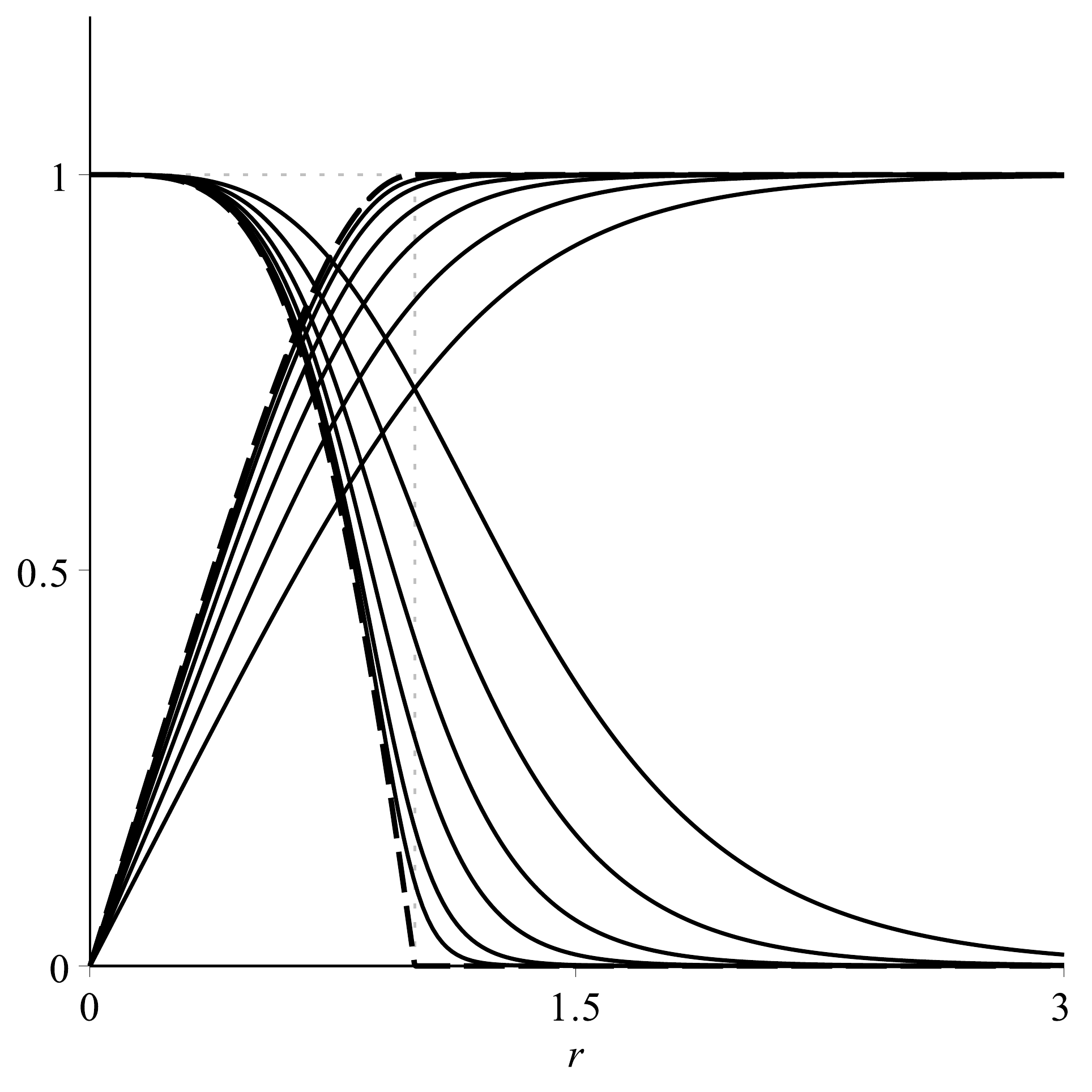}} 
\caption{The solutions $a(r)$ and $g(r)$ for $n=1$. We first consider $l=1$ and then increase it to larger and larger values. The dashed lines stand for the compact limit, given by Eqs.~\eqref{solhc} with $n=1$.} 
\label{fig11}
\end{figure}
%%%%%%%%%%%%%%%%%%%%%%%%%%%%%%

The Eqs.~\eqref{25} now become
\begin{equation}\label{eqahc}
g'= \pm\frac{ag}{r}, \quad \quad
\frac{a^\prime}{r}= \mp 3 g^2(1-g^{2l}).
\end{equation}
Near the origin, we take $a(r)\approx n+a_0(r)$ and $g(r) \approx g_0(r)$; thus, up to first-order in $a_0(r)$ and $g_0(r)$ one gets
\ben
a_0(r) = 0, \quad \quad
g_0(r) = \alpha r^n,
\een
where $\alpha$ is an integration constant. Asymptotically, we take $a(r)\approx a_{asy}(r)$ and $g(r) \approx 1+g_{asy}(r)$ to get
\bes
\ben
a_{asy}(r) &=& \sqrt{6l}\beta r K_1\left(\sqrt{6l}r\right), \\
g_{asy}(r) &=& -\beta K_0\left(\sqrt{6l}r\right),
\een 
\ees
where $K_\nu(x)$ is the modified Bessel function of the second kind and $\beta$ is an integration constant. For a general $l$ we focus to solve
Eqs.~\eqref{eqahc} numerically. However, for $n$ positive and for very large $l$, it is possible to show that the model supports the compact solutions
\bes\label{solhc}
\ben
a_c(r) &=&
\begin{cases}
\frac{(n+2)\left((n(n+2))^{n+1} -(3r^2)^{n+1} \right)}{(n+2)^2\left(n(n+2)\right)^n+(3r^2)^{n+1}},\,\,&r\leq r_c\\
0, \,\, & r>r_c,
\end{cases} \nonumber \\
\\
g_c(r) &=&
\begin{cases}
\frac{2(n+1)(n+2)\left(3n(n+2)\right)^{n/2} r^n}{(n+2)^2\left(n(n+2)\right)^n+ (3r^2)^{n+1}},\,\,\,&r\leq r_c\\
1, \,\,\, & r>r_c,
\end{cases} \nonumber \\
\een
\ees
where $r_c=\sqrt{n(n+2)/3}$ is the radius of the compact solutions. The discussion about the regularity of the solutions is similar to the previous one, so we omit it here. In Fig.~\ref{fig11}, we display the solutions for $n=1$ and for several values of $l$.

We then focus on the energy density and magnetic field. As it can be checked, the expression for the energy density in the compact limit is cumbersome, so we omit it here. However, the magnetic field gets the form
\be\label{bhc}
B_c(r)=
\begin{cases}
\frac{12\left((n+1)(n+2)\right)^2 \left(3n(n+2)r^2\right)^n}{\left((n(n+2))^n (n+2)^2 + (3r^2)^{n+1}\right)^2},\,\,\,&r\leq r_c \\
0, \,\,\, & r>r_c.
\end{cases}
\ee
In Fig.~\eqref{fig12} we display how the energy density and magnetic field behave for $n=1$ and for several values of $l$. As in the previous model, the discontinuity of the magnetic field in the compact limit induces no problem here too, since it is also controlled by the presence of the generalized magnetic permeability.

%%%%%%%%%%%%%%%%%%%%%%%%%%%%%%%%%%%%%%%%%%%%%%%%%%%
\section{Comments and conclusions}

In this work we studied the presence of vortices in a generalized Maxwell-Higgs model. The main idea was to generalize the Maxwell-Higgs model in a way such that we could find first-order differential equations and explore the BPS solutions. To do this, we have changed the Maxwell term, adding to it the factor $G(|\phi|)$, which seems to model a generalized magnetic permeability. As one knows, this modification leads to effective planar field theories that present vortex solutions which somehow describe the vortices of the models with standard Maxwell and Chern-Simons dynamics.

Despite the change in the Maxwell term, one could write a first-order framework and find vortices which are similar to the vortices of models with         standard Maxwell and Chern-Simons dynamics. Moreover, we could modify the function $G(|\phi|)$ to describe different models, with the solutions having distinct profiles, as we studied in Sec.~\ref{sec3}. We then used this fact to propose other models in Sec.~\ref{sec4}, with focus on the possibility to shrink the solutions to a compact interval of the radial coordinate, in a way similar to the case of compact kinks investigated before in \cite{CK}. We then studied two distinct models, one similar to the model with standard Maxwell dynamics, and the other bringing resemblance with the Chern-Simons dynamics. 

%%%%%%%%%%%%%%%%%%%%%%%%%%%%%%%
\begin{figure}[t!]
{\includegraphics[width=4.2cm]{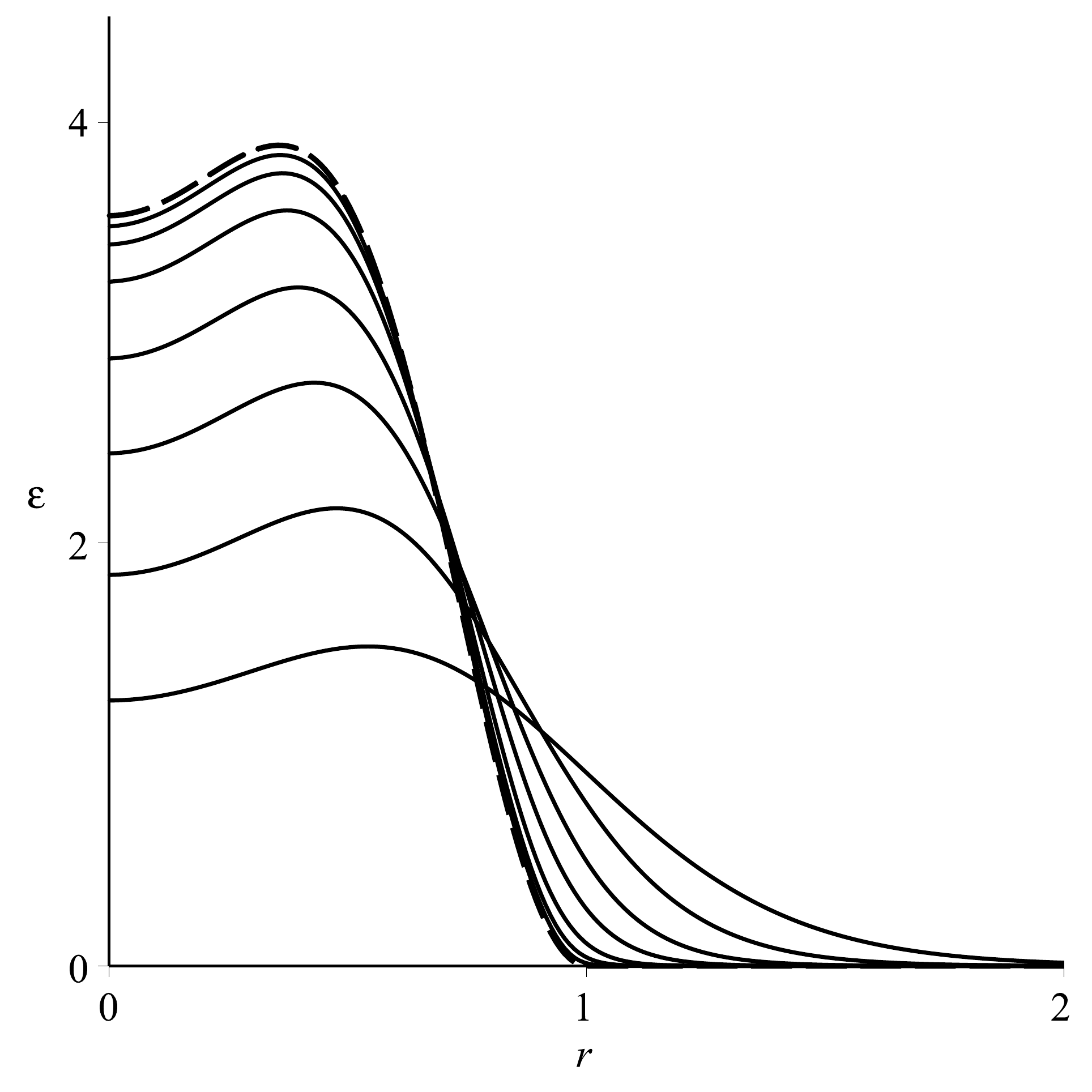}}
{\includegraphics[width=4.2cm]{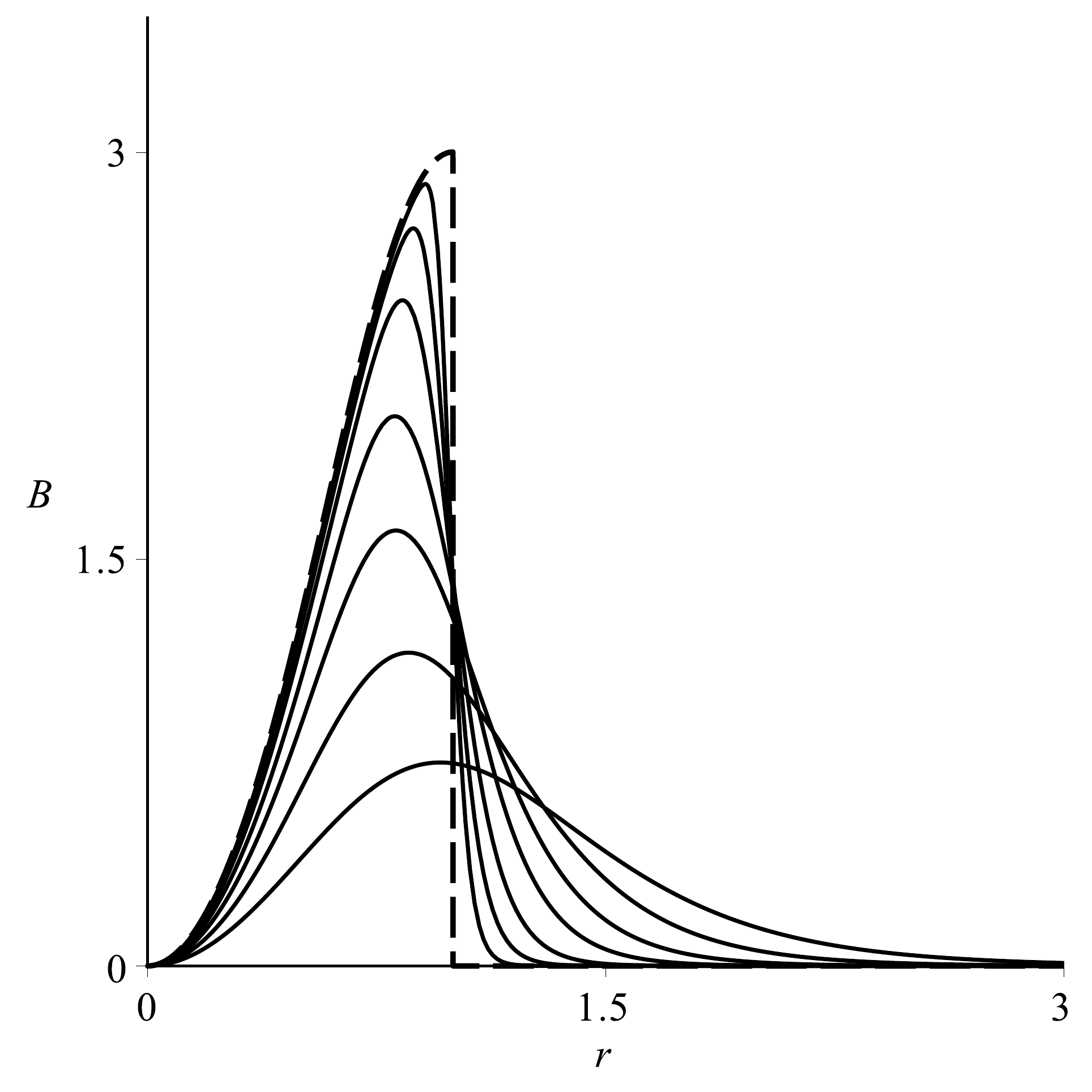}} 
\caption{The energy density (left), and the magnetic field (right) for $n=1$. We first consider $l=1$ and then increase it to larger and larger values. The dashed lines represent the compact limit, in which the magnetic field is given by Eq.~\eqref{bhc} with $n=1$.} \label{fig12}
\end{figure}
%%%%%%%%%%%%%%%%%%%%%%%%%%%%%%%%%%%%%%%%%%%%%%%%%%%

The results indicated the presence of the compact behavior, with the vortices shrinking to a compact interval, with the energy density and magnetic field vanishing outside the compact interval. The compact behavior appears very clearly in Figs.~\ref{fig8} and \ref{fig9} for the model \eqref{Vc}, and in Figs.~\ref{fig11} and \ref{fig12}, for the model \eqref{Vhc}. The two models are different from each other: the first one, described by the potential \eqref{Vc} is similar to the standard Maxwell-Higgs model, and the other, with potential \eqref{Vhc}, resembles the model with Chern-Simons dynamics.

We identified a new behavior, a compact behavior for the vortices that appear in the models studied in Sec.~\ref{sec4}. This seems to be of current interest, and we hope that the above results will stimulate further research in the area, especially on the main characteristics of the solutions, and in the construction of new models. Interesting issues concern extending the current results to other topological structures, in particular to monopoles and skyrmions. The case of skyrmions is of practical interest, and the study of compact skyrmions can be used to describe new spin textures in high energy physics \cite{adam} and in magnetic materials \cite{MS1,MS2,sky}. Research in this direction is now under development, and we hope to report on them in the near future.

%%%%%%%%%%%%%%%%%%%%%%%%%%%%%%%%%%%%%%%%%%%%%%%%%%%
\section*{Acknowledgments}
This work is partially supported by CNPq, Brazil. DB acknowledges support from projects CNPq:455931/2014-3 and CNPq:06614/2014-6, LL acknowledges support from projects CNPq:307111/2013-0 and CNPq:447643/2014-2, MAM thanks support from project CNPq:140735/2015-1, and RM thanks support from projects CNPq:508177/2010-3 and CNPq:455619/2014-0.

%%%%%%%%%%%%%%%%%%%%%%%%%%%%%%%%%%%%%%%%%%%%%%%%%%%

%%%%%%%%%%%%%%%%%%%%%%%%%%%%%%%%%%%%%%%%%%%%%%%%%%%
\end{document}